\documentclass[12pt]{article}
\usepackage{amscd}
\usepackage{bbm}
\usepackage{mathrsfs}
\usepackage{amssymb}
\usepackage{graphics}
\usepackage{graphicx}
\usepackage{amsfonts}
\usepackage{amsmath}
\usepackage{caption}
\usepackage{titlesec}
\usepackage{subcaption}
\usepackage{float}
\usepackage{booktabs,multirow}
\usepackage{array}
\usepackage[numbers,sort&compress]{natbib}
\titleformat{\section}{\large\bfseries}{\thesection. }{0em}{}
\titleformat{\subsection}{\normalsize\bfseries}{\thesubsection. }{0em}{}
\titleformat{\subsubsection}{\small\bfseries}{\thesubsubsection. }{0em}{}
\numberwithin{equation}{section}
\begin{document}
\date{}
\title{Analytical and numerical studies for integrable and non-integrable fractional discrete modified Korteweg-de Vries hierarchies}

\author{Qin-Ling Liu, Rui Guo$ \thanks{Corresponding author,
gr81@sina.com}$, Ya-Hui Huang, Xin Li \
\\
\\{\em
School of Mathematics, Taiyuan University  of} \\
{\em Technology, Taiyuan 030024, China} } \maketitle

\begin{abstract}

Under investigation in this paper is the fractional integrable and non-integrable discrete modified Korteweg-de Vries hierarchies. The linear dispersion relations, completeness relations, inverse scattering transform, and fractional soliton solutions of the fractional integrable discrete modified Korteweg-de Vries hierarchy will be explored. The inverse scattering problem will be solved accurately by using Gel'fand-Levitan-Marchenko (GLM) equations and Riemann-Hilbert (RH) problem.  The peak velocity of fractional soliton solutions will be analyzed. The numerical solutions of the non-integrable fractional averaged discrete modified Korteweg-de Vries equation which has a simpler form than the integrable one will be obtained by a split-step fourier method.

\vspace{7mm}\noindent\emph{Keywords}: The fractional integrable and non-integrable discrete modified Korteweg-de Vries hierarchies; The linear dispersion relations; The inverse scattering transform; The numerical solution; Fractional soliton solution;
\end{abstract}

\newpage
\section{Introduction}
\hspace{0.7cm}Nonlinear dynamics is a subject that studies the qualitative and quantitative changes of various motion states of nonlinear dynamic systems, which involves many fields of natural science and social science~\cite{g1,g2,g3}. As a foundation of nonlinear dynamics, nonlinear integrable equations are widely used in complex physical phenomena in the field of physics, such as fluid mechanics~\cite{g4}, solid state physics~\cite{g5}, plasma waves~\cite{g6}, nonlinear optics~\cite{g7}, etc. In order to obtain the explicit solution of nonlinear integrable equations, many methods have been developed, including inverse scattering transform (IST)~\cite{g8}, Darboux transformation (DT)~\cite{g9}, Hirota's bilinear method~\cite{g10} and Jacobi elliptic function expansion method~\cite{g11}, among which the IST is a systematic approach to solving. This method has a rigorous physical background and mathematical rigor, and it is possible to solve the entire family of isospectral development equations linked to the same spectral problem with multiple soliton solutions~\cite{g12,g13}.

The concept of fractional calculus appeared almost simultaneously with that of integer calculus. The origin of fractional calculus can be traced back to 1695, with contributions from Leibinz, the founder of classical calculus, and the great mathematicians Euler, Liouville, Riemann, Cauchy, Weyl, and others~\cite{g14,g15}. The most important advantage of fractional differential equations over integer differential equations is that they can better simulate physical and dynamic system processes in nature~\cite{g16,g17,g18}. For example, when the diffusion propagation rate of a particle is inconsistent with the classical Brownian motion model, the fractional order derivative plays a very crucial role in modeling the anomalous motion of this particle, and replacing the second-order derivative of spatial diffusion with the fractional order derivative in the diffusion space model will lead to stronger diffusion~\cite{g19,g20,g21}. Fractional order derivatives are so nonlocal that they are well suited for describing some phenomena with heredity and memory. So far, fractional calculus has been successfully applied in many fields such as high-energy physics~\cite{g22}, anomalous diffusion~\cite{g23}, viscoelastic materials~\cite{g24}, solid mechanics~\cite{g25}, and finance~\cite{g26}.

In 2022, as a combination of fractional calculus and nonlinear integrable equation, a new kind of fractional equation was proposed, namely fractional nonlinear Schr\"{o}dinger (fNLS) equation and fractional Korteweg-deVries (fKdV) equation~\cite{g27}. These fractional equations are integrable in the sense of IST which can predict superdispersive transport in fractional media. After that this fractional equation can be applied to the whole Ablowitz-Kaup-Newell-Segur (AKNS) system, i.e., integrable fractional modified Korteweg-de Vries (fmKdV), sine-Gordon (fsineG), and sinh-Gordon (fsinhG) equations were studied~\cite{g28}. The higher-order integrable fractional mKdV equation and higher-order integrable fractional NLS equation were discussed~\cite{g29,g30}. The fractional integrable coupled Hirota equation with $3\times3$ Lax pair was investigated~\cite{g31}, and the integrable fractional n-component coupled nonlinear Schr\"{o}dinger model was also presented~\cite{g32}. Recently, the fractional integrable equation has been extended from the AKNS system to the Kaup-Newell (KN) system, which described three kinds of fractional derivative type NLS equations~\cite{g33} and fractional Fokas-Lenells equation~\cite{g34}. Ref. [35] has extended from a continuous fractional equation to a discrete fractional NLS equation. To date, to our knowledge, the discrete fractional integrable mKdV hierarchy has not been obtained. In this paper, discrete equations are semi-discrete, that is, discrete in space and continuous in time. The discrete fractional integrable mKdV hierarchy will be developed which requires three key steps: linear dispersion relations, completeness relations, and IST.

The research on many problems in natural science can be roughly divided into two categories: qualitative research and quantitative research. In quantitative research, it can be subdivided into numerical approximate research and accurate structural research~\cite{g36}. The above-mentioned research on the fractional integrable equations belongs to the accurate structural research. Hereafter, we study the numerical approximate research of another discrete fractional mKdV equation, i.e., fractional averaged discrete mKdV (fAdmKdV) equation, which is non-integrable in the sense of IST, but simpler than the integrable one in fractional structure. One of the numerical methods is the split-step method, which decomposes the original problem into linear and nonlinear subproblems, and then approaches the solution of the original problem by solving the subproblems in turn by various methods~\cite{g40}. Various versions of the split-step method have been used in many classical equations, such as NLS equation~\cite{g37}, KdV equation~\cite{g38}, mKdV equation~\cite{g39}, complex mKdV equation~\cite{g40}, generalized NLS equation~\cite{g41}, space fractional variable-coefficient KdV-mKdV equation~\cite{g42}. What's more, it has been applied to the fractional averaged discrete NLS equation~\cite{g35}.

The layout of the paper is as follows. In Section 2 the explicit forms and linear dispersive relations of the fractional integrable discrete modified Korteweg-de Vries hierarchy will be obtained. Next, in Section 3, we will pay attention to a completeness relation for the squared eigenfunctions of the fractional integrable discrete modified Korteweg-de Vries hierarchy. In addition, the long Section 4 will be devoted to analysis of the inverse scattering problem, and the fractional single-poles soliton solutions for the reflectionless potentials will be gained. Finally, the numerical solution of the non-integrable fractional averaged discrete modified Korteweg-de Vries equation which has a simpler structure than the integrable one will be obtained in Section 5.

\section{Fractional integrable discrete mKdV hierarchy}
\hspace{0.6cm}In this section, we present a family of fractional integrable discrete mKdV equations and give the linear dispersion relation.

Based on the results in Refs.~\cite{g43} and~\cite{g44}, the integrable discrete mKdV hierarchy can be characterized by the dispersion relation of its associated linearized version and a discrete sum-difference operator.
\begin{equation}
\frac{d}{d\tau}\left( \begin{array}{c}
	Q_n\\
	R_n\\
\end{array} \right) +\mathscr{N}_n\left( \mathcal{L} \right) \left( \begin{array}{c}
	Q_n\\
	-R_n\\
\end{array} \right) =0,
\end{equation}
where $\mathcal{L}$ is a sum-difference operator given by
\begin{equation}
\mathcal{L}=\left( \begin{matrix}
	E_{n}^{+}+Q_{n+1}\varDelta _n\mathscr{J}_{n+1}\left( R_j \right) +Q_n\mathscr{S}_{n+1}^{+}\left( R_j \right)&		Q_{n+1}\varDelta _n\mathscr{J}_{n+1}\left( Q_j \right) +Q_n\mathscr{S}_{n+1}^{-}\left( Q_j \right)\\
	R_{n-1}\varDelta _n\mathscr{J}_n\left( R_j \right) -R_n\mathscr{S}_{n}^{+}\left( R_j \right)&		E_{n}^{-}-R_{n-1}\varDelta _n\mathscr{J}_n\left( -Q_j \right) -R_n\mathscr{S}_{n}^{-}\left( Q_j \right)\\
\end{matrix} \right) ,
\end{equation}
where $E_{n}^{\pm}u_n=u_{n\pm 1}$ and $\varDelta_n=1-R_nQ_n$. The operators $\mathscr{J}_n$ and $\mathscr{S}_{n}^{\pm}$ are defined by
\begin{subequations}
\begin{align}
\mathscr{J}_n\left( x_j \right) y_n &\equiv \sum_{j=n}^{\infty}{\frac{x_jy_j}{\varDelta_j}},\\
\mathscr{S}_{n}^{\pm}\left( x_j \right) y_n &\equiv \sum_{j=n}^{\infty}{x_jy_{j\pm 1}}.
\end{align}
\end{subequations}
The inverse of the sum-difference operator is
\begin{equation}
\mathcal{L}^{-1}=\left( \begin{matrix}
	E_{n}^{-}-Q_{n-1}\varDelta _n\mathscr{J}_n\left( -R_j \right) -Q_n\mathscr{S}_{n}^{-}\left( R_j \right)&		Q_{n-1}\varDelta _n\mathscr{J}_n\left( Q_j \right) -Q_n\mathscr{S}_{n}^{+}\left( Q_j \right)\\
	R_{n+1}\varDelta _n\mathscr{J}_{n+1}\left( -R_j \right) +R_n\mathscr{S}_{n+1}^{-}\left( R_j \right)&		E_{n}^{+}-R_{n+1}\varDelta _n\mathscr{J}_{n+1}\left( Q_j \right) +R_n\mathscr{S}_{n+1}^{+}\left( Q_j \right)\\
\end{matrix} \right).
\end{equation}
$\mathscr{N}_n\left( \mathcal{L} \right) =\sum_{j=1}^n{\alpha _j\left( \left( \mathcal{L}^{-1} \right) ^j-\mathcal{L}^j \right)}$ is connected with the linearized dispersion relation of the integrable discrete mKdV hierarchy with $\alpha _j\in \mathbb{R}$. In particular, when $n=1,\ \alpha _1=1$ and $\mathscr{N}_1\left( \mathcal{L} \right) =\mathcal{L}^{-1}-\mathcal{L}$, one yields
\begin{equation}
Q_{n,\tau}+\left( 1\pm Q_{n}^{2} \right) \left( Q_{n-1}-Q_{n+1} \right) =0.
\end{equation}
When $n=2,\ \alpha _1=-1,\ \alpha _2=\frac{1}{2}$ and $\mathscr{N}_2\left( \mathcal{L} \right) =1/\left(2\mathcal{L}^{-2}\right) -\mathcal{L}^{-1} +\mathcal{L}-\mathcal{L}^{2} /2$, one has
\begin{equation}
Q_{n,\tau}+\left( 1\pm Q_{n}^{2} \right) \left( Q_{n+1}-Q_{n-1}-\frac{1}{2}Q_{n+2}\left( 1\pm Q_{n+1}^{2} \right) +\frac{1}{2}Q_{n-2}\left( 1\pm Q_{n-1}^{2} \right) \pm \right. \nonumber
\end{equation}
\begin{equation}
 \left.\frac{1}{2}Q_n\left( Q_{n+1}^{2}-Q_{n-1}^{2} \right) \right)=0.
\end{equation}

The linearization of the integrable discrete mKdV hierarchy is
\begin{equation}
\frac{d}{d\tau}\left( \begin{array}{c}
	Q_n\\
	R_n\\
\end{array} \right) +\left( \begin{matrix}
	\mathscr{N}_n\left( E_{n}^{+} \right)&		0\\
	0&		\mathscr{N}_n\left( E_{n}^{-} \right)\\
\end{matrix} \right) \left( \begin{array}{c}
	Q_n\\
	-R_n\\
\end{array} \right) =0.
\end{equation}
If we let $Q_n=z^{2n}e^{\omega_n \left( z \right) \tau}$ into Eq. (2.7), we obtain
\begin{equation}
\mathscr{N}_n\left( z^2 \right) =-\omega_n \left( z \right),
\end{equation}
where $\omega_n \left( z \right)$ is the dispersion relation of the linearized version of the integrable discrete mKdV hierarchy.

In order to define the fractional integrable discrete mKdV hierarchy, we put the dispersion relation $\omega_n \left( z \right) \sim -\left( \sum_{j=1}^n{\alpha _j\left( z^{-2j}-z^{2j} \right)} \right) ^{1+\epsilon}$, it can be directly given that $\mathscr{N}_n\left( \mathcal{L} \right) \sim \left( \sum_{j=1}^n{\alpha _j\left( \left( \mathcal{L}^{-1} \right) ^j-\mathcal{L}^j \right)} \right) ^{1+\epsilon}$ by using Eq. (2.8), therefore, the fractional integrable discrete mKdV hierarchy is
\begin{equation}
\frac{d}{d\tau}\left( \begin{array}{c}
	Q_n\\
	R_n\\
\end{array} \right) +\left( \sum_{j=1}^n{\alpha _j\left( \left( \mathcal{L}^{-1} \right) ^j-\mathcal{L}^j \right)} \right) ^{1+\epsilon}\left( \begin{array}{c}
	Q_n\\
	-R_n\\
\end{array} \right) =0.
\end{equation}
Specifically, as $n=1$, one yields
\begin{equation}
\frac{d}{d\tau}\left( \begin{array}{c}
	Q_n\\
	R_n\\
\end{array} \right) +\left( \mathcal{L}^{-1}-\mathcal{L} \right) ^{1+\epsilon}\left( \begin{array}{c}
	Q_n\\
	-R_n\\
\end{array} \right) =0,
\end{equation}
as $n=2$, one can get
\begin{equation}
\frac{d}{d\tau}\left( \begin{array}{c}
	Q_n\\
	R_n\\
\end{array} \right) +\left( \frac{1}{2\mathcal{L}^{-2}}-\mathcal{L}^{-1}+\mathcal{L}-\frac{\mathcal{L}^2}{2} \right) ^{1+\epsilon}\left( \begin{array}{c}
	Q_n\\
	-R_n\\
\end{array} \right) =0.
\end{equation}
It can be seen intuitively that the limits of Eqs. (2.10) and (2.11) as $\epsilon\rightarrow0$ are Eqs. (2.5) and (2.6) respectively. In the following, we mainly solve the above two formulas. In fact, one can also consider the fractional high-order integrable discrete mKdV equations that are generated from Eq. (2.9) in the same way.

\section{Completeness of squared eigenfunctions and fractional operators}
\hspace{0.6cm}Let us introduce the systems $\Psi$, $\bar{\Psi}$ of ``squared'' solutions of the discrete scattering problem by
\begin{subequations}
\begin{align}
\Psi_n \left( z \right) =\left( \begin{matrix}
	c_n\psi _{n}^{\left( 1 \right)}\left( z \right) \psi _{n+1}^{\left( 1 \right)}\left( z \right) ,&		c_n\psi _{n}^{\left( 2 \right)}\left( z \right) \psi _{n+1}^{\left( 2 \right)}\left( z \right)\\
\end{matrix} \right) ^T, \\
\bar{\Psi}_n \left( z \right) =\left( \begin{matrix}
	c_n\bar{\psi}_{n}^{\left( 1 \right)}\left( z \right) \bar{\psi}_{n+1}^{\left( 1 \right)}\left( z \right) ,&		c_n\bar{\psi}_{n}^{\left( 2 \right)}\left( z \right) \bar{\psi}_{n+1}^{\left( 2 \right)}\left( z \right)\\
\end{matrix} \right) ^T,
\end{align}
\end{subequations}
these systems are also eigenfunctions of the sum-difference operator $\mathcal{L}$, and the eigenfunctions of its adjoint operator $\widetilde{\mathcal{L}}$ are as follows
\begin{subequations}
\begin{align}
\widetilde{\Psi}_n \left( z \right) =\left( \begin{matrix}
	c_n\phi _{n}^{\left( 2 \right)}\left( z \right) \phi _{n+1}^{\left( 2 \right)}\left( z \right) ,&		-c_n\phi _{n}^{\left( 1 \right)}\left( z \right) \phi _{n+1}^{\left( 1 \right)}\left( z \right)\\
\end{matrix} \right) ^T, \\
\widetilde{\bar{\Psi}}_n \left( z \right) =\left( \begin{matrix}
	c_n\bar{\phi}_{n}^{\left( 2 \right)}\left( z \right) \bar{\phi}_{n+1}^{\left( 2 \right)}\left( z \right) ,&		-c_n\bar{\phi}_{n}^{\left( 1 \right)}\left( z \right) \bar{\phi}_{n+1}^{\left( 1 \right)}\left( z \right)\\
\end{matrix} \right) ^T,
\end{align}
\end{subequations}
where $\phi\left( z \right), \bar{\phi}\left( z \right), \psi\left( z \right), \bar{\psi}\left( z \right), c_n$ will be defined in the next section.

The squared eigenfunctions of the operator $\mathscr{N}_n\left( \mathcal{L} \right)$ and $\mathscr{N}_n\left( \widetilde{\mathcal{L}} \right)$ and eigenvalue $z$ satisfy
\begin{subequations}
\begin{align}
\mathscr{N}_n\left( \mathcal{L} \right) \Psi _n=\mathscr{N}_n\left( z^2 \right) \Psi _n,\ \ \ \ \mathscr{N}_n\left( \mathcal{L} \right) \Psi _n=\mathscr{N}_n\left( z^2 \right) \Psi _n, \\
\mathscr{N}_n\left( \widetilde{\mathcal{L}} \right) \widetilde{\Psi} _n=\mathscr{N}_n\left( z^2 \right) \widetilde{\Psi} _n,\ \ \ \mathscr{N}_n\left( \widetilde{\mathcal{L}} \right) \widetilde{\bar{\Psi}} _n=\mathscr{N}_n\left( z^2 \right) \widetilde{\bar{\Psi}} _n.
\end{align}
\end{subequations}
Because the squared eigenfunctions are complete, and applying the operator $\mathscr{N}_n\left( \mathcal{L} \right)$  to arbitrary discrete function $\chi _n=\left( \begin{matrix}
	\chi _{n}^{\left( 1 \right)},&		\chi _{n}^{\left( 2 \right)}\\
\end{matrix} \right)^T $, one has
\begin{equation}
\mathscr{N}_n\left( \mathcal{L} \right) \chi _n=\frac{i}{2\pi}\sum_{j=1}^2{\oint_{S^{\left( j \right)}}{\frac{dz}{z}}}\mathscr{N}_n\left( z^2 \right) \sum_{m=-\infty}^{\infty}{\mathcal{G}_{n,m}^{\left( j \right)}\left( z \right) \chi _m},
\end{equation}
where $S^{\left( j \right)}\left( j=1,2 \right)$ are the negatively oriented circles, where the radius of $S^{\left( 1 \right)}$ is greater than 1 and the radius of $S^{\left( 2 \right)}$ is less than 1, and,
\begin{subequations}
\begin{align}
\mathcal{G}_{n,m}^{\left( 1 \right)}\left( z \right) &=\frac{\varPsi _n\left( z \right) \widetilde{\varPsi }_m\left( z \right) ^T}{\left( 1-Q_{n}^{2} \right) a^2\left( z \right)},\\
\mathcal{G}_{n,m}^{\left( 2 \right)}\left( z \right) &=-\frac{\varPsi _n\left( z \right) \widetilde{\varPsi }_m\left( z \right) ^T}{\left( 1-Q_{n}^{2} \right) \bar{a}^2\left( z \right)}.
\end{align}
\end{subequations}
By choosing the arbitrary discrete function of the above formula as $\left(\begin{matrix}
	Q_n,&		-R_n\\
\end{matrix} \right)^T $, and using Eq. (2.1),  the fractional integrable discrete mKdV hierarchy is given by
\begin{equation}
\frac{dQ_n}{dt}=\frac{i}{2\pi}\sum_{j=1}^2{\oint_{S^{\left( j \right)}}{\frac{dz}{z}}}\mathscr{N}_n\left( z^2 \right) \sum_{m=-\infty}^{\infty}{\mathfrak{g}_{n,m}^{\left( j \right)}\left( z \right)},
\end{equation}
with
\begin{subequations}
\begin{align}
\mathfrak{g}_{n,m}^{\left( 1 \right)}\left( z \right) &=-\frac{c_nc_m}{\left( 1-Q_{n}^{2} \right) a^2\left( z \right)}\psi _{n}^{\left( 1 \right)}\left( z \right) \psi _{n+1}^{\left( 1 \right)}\left( z \right) \left( \phi _{m}^{\left( 1 \right)}\left( z \right) \phi _{m+1}^{\left( 1 \right)}\left( z \right) Q_m+\phi _{m}^{\left( 2 \right)}\left( z \right) \phi _{m+1}^{\left( 2 \right)}\left( z \right) R_m \right),\\
\mathfrak{g}_{n,m}^{\left( 2 \right)}\left( z \right) &=\frac{c_nc_m}{\left( 1-Q_{n}^{2} \right) a^{*2}\left( 1/z^* \right)}\left( \psi _{n}^{\left( 2 \right)}\left( 1/z^* \right) \psi _{n+1}^{\left( 2 \right)}\left( 1/z^* \right) \right) ^*\times \\  \nonumber
&\ \ \ \ \ \left( \phi _{m}^{\left( 1 \right)}\left( 1/z^* \right) \phi _{m+1}^{\left( 1 \right)}\left( 1/z^* \right) R_m+\phi _{m}^{\left( 2 \right)}\left( 1/z^* \right) \phi _{m+1}^{\left( 2 \right)}\left( 1/z^* \right) Q_m \right) ^*.
\end{align}
\end{subequations}

\section{Inverse scattering transform}
\hspace{0.6cm}In this section, we solve the fractional integrable discrete modified Korteweg-de Vries equations with the IST. The IST has three distinct steps: direct scattering, time evolution, and inverse scattering.
\subsection{Direct scattering}
\hspace{0.6cm}The discrete scattering problem of the fractional integrable discrete mKdV equation is
\begin{equation}\label{f4.1}
v_{n+1}=\left( \begin{matrix}
	z&		Q_n\\
	R_n&		z ^{-1}\\
\end{matrix} \right) v_n.
\end{equation}
With sufficient decay and smoothness of the potential function, the discrete scattering problem~(\ref{f4.1}) is asymptotic to
\begin{equation}\label{f4.2}
v_{n+1}\sim\left( \begin{matrix}
	z&		0\\
	0&		z ^{-1}\\
\end{matrix} \right) v_n.
\end{equation}
Eigenfunctions of the discrete scattering system, namely, the solutions of Eq.~(\ref{f4.1}), satisfy the boundary conditions
\begin{subequations}
\begin{align}
\phi _n\left( z \right) \sim z^n\left( \begin{array}{c}
	1\\
	0\\
\end{array} \right) ,\ \bar{\phi}_n\left( z \right) \sim z^{-n}\left( \begin{array}{c}
	0\\
	1\\
\end{array} \right) ,\ as\ n\rightarrow -\infty, \\
\psi _n\left( z \right) \sim z^{-n}\left( \begin{array}{c}
	0\\
	1\\
\end{array} \right) ,\,\,\bar{\psi}_n\left( z \right) \sim z^n\left( \begin{array}{c}
	1\\
	0\\
\end{array} \right) ,\,\,as\,\,n\rightarrow +\infty.
\end{align}
\end{subequations}
Therefore, we define the four Jost functions as follows
\begin{subequations}
\begin{align}
M_n\left( z \right) =z^{-n}\phi _n\left( z \right) ,\ \bar{M}_n\left( z \right) =z^n\bar{\phi}_n\left( z \right) ,\,\, \\
N_n\left( z \right) =z^n\psi _n\left( z \right) ,\,\,\bar{N}_n\left( z \right) =z^{-n}\bar{\psi}_n\left( z \right) ,\,\,
\end{align}
\end{subequations}
and the Jost functions satisfy the asymptotic conditions
\begin{subequations}
\begin{align}
M_n\left( z \right) \sim \left( \begin{array}{c}
	1\\
	0\\
\end{array} \right) ,\,\,\bar{M}_n\left( z \right) \sim \left( \begin{array}{c}
	0\\
	1\\
\end{array} \right) ,\,\,as\,\,n\rightarrow -\infty, \\
N_n\left( z \right) \sim \left( \begin{array}{c}
	0\\
	1\\
\end{array} \right) ,\,\bar{N}_n\left( z \right) \sim \left( \begin{array}{c}
	1\\
	0\\
\end{array} \right) ,\,\,as\,\,n\rightarrow +\infty.
\end{align}
\end{subequations}
The discrete scattering problem~(\ref{f4.1}) is a second-order difference equation, so there are at most two linearly independent solutions for any fixed value of $z$. Because the eigenfunctions $\psi _n$ and $\bar{\psi}_n$ are linearly independent, the eigenfunctions $\phi_n$ and $\bar{\phi}_n$ can be expressed as
\begin{subequations}\label{f4.6}
\begin{align}
\phi _n\left( z \right) =b\left( z \right) \psi _n\left( z \right) +a\left( z \right) \bar{\psi}_n\left( z \right),\\
\bar{\phi}_n\left( z \right) =\bar{a}\left( z \right) \psi _n\left( z \right) +\bar{b}\left( z \right) \bar{\psi}_n\left( z \right).
\end{align}
\end{subequations}
The scattering data can be represented in terms of Wronskians of the eigenfunctions,
\begin{subequations}
\begin{align}
a\left( z \right) &=c_nWr\left( \begin{matrix}
	\phi _n\left( z \right) ,&		\,\psi _n\left( z \right)\\
\end{matrix} \right) =c_nWr\left( \begin{matrix}
	M_n\left( z \right) ,&		\,N_n\left( z \right)\\
\end{matrix} \right),\\
\bar{a}\left( z \right) &=c_nWr\left( \begin{matrix}
	\bar{\psi}_n\left( z \right) ,&		\bar{\phi}_n\left( z \right)\\
\end{matrix} \right) =c_nWr\left( \begin{matrix}
	\bar{N}_n\left( z \right) ,&		\bar{M}_n\left( z \right)\\
\end{matrix} \right),\\
b\left( z \right) &=c_nWr\left( \begin{matrix}
	\bar{\psi}_n\left( z \right) ,&		\phi _n\left( z \right)\\
\end{matrix} \right) =z ^{2n}c_nWr\left( \begin{matrix}
	\bar{N}_n\left( z \right) ,&		M_n\left( z \right)\\
\end{matrix} \right),\\
\bar{b}\left( z \right) &=c_nWr\left( \begin{matrix}
	\bar{\phi}_n\left( z \right) ,&		\psi _n\left( z \right)\\
\end{matrix} \right) =z ^{-2n}c_nWr\left( \begin{matrix}
	\bar{M}_n\left( z \right) ,&		N_n\left( z \right)\\
\end{matrix} \right),
\end{align}
\end{subequations}
with $c_n=\prod_{k=n}^{+\infty}{\left( 1-Q_{n}^{2} \right)}$.

Note that Eqs.~(\ref{f4.6}a) and~(\ref{f4.6}b) can be written as
\begin{subequations}
\begin{align}
\mu _n\left( z \right) &=z ^{-2n}\rho \left( z \right) N_n\left( z \right) +\bar{N}_n\left( z \right),\\
\bar{\mu}_n\left( z \right) &=N_n\left( z \right) +z ^{2n}\bar{\rho}\left( z \right) \bar{N}_n\left( z \right),
\end{align}
\end{subequations}
where $\mu _n\left( z \right) =\frac{M_n\left( z \right)}{a\left( z \right)},\ \bar{\mu}_n\left( z \right) =\frac{\bar{M}_n\left( z \right)}{\bar{a}\left( z \right)}$ and the reflection coefficients $\rho \left( z \right) =\frac{b\left( z \right)}{a\left( z \right)}, \bar{\rho}\left( z \right) =\frac{\bar{b}\left( z \right)}{\bar{a}\left( z \right)}$.

Suppose that $a\left( z \right)$ has $J$ simple zeros denoted by $z_j(j=1,2,\cdots,J)$, and $\bar{a}\left( z \right)$ has $\bar{J}$ simple zeros denoted by $\bar{z}_\epsilon(\epsilon=1,2,\cdots,\bar{J})$. Indeed, $\phi _n\left( z _j \right),\psi _n\left( z _j \right)$ and $\bar{\phi}_n\left( \bar{z}_{\epsilon} \right),\bar{\psi}_n\left( \bar{z}_{\varepsilon} \right)$ are linearly dependent, that is,
\begin{equation}\label{f4.9}
\phi _n\left( z _j \right) =b_j\psi _n\left( z _j \right) ,\ \ \bar{\phi}_n\left( \bar{z}_{\epsilon} \right) =\bar{b}_{\epsilon}\bar{\psi}_n\left( \bar{z}_{\varepsilon} \right).
\end{equation}
In terms of the Jost functions, Eq.~(\ref{f4.9}) can be written as
\begin{equation}
M_n\left( z _j \right) =b_jz _j^{-2n}N_n\left( z _j \right) ,\,\,\,\,\bar{M}_n\left( \bar{z}_{\epsilon} \right) =\bar{b}_{\epsilon}\bar{z}_{\epsilon}^{2n}\bar{N}_n\left( \bar{z}_{\varepsilon} \right).
\end{equation}
We define the norming constants associated with the eigenvalues $z_j$ and $\bar{z}_{\epsilon}$ by
\begin{equation}
C_j=\frac{b_j}{a'\left( z _j \right)} ,\ \  \bar{C}_{\epsilon}=\frac{\bar{b}_{\epsilon}}{\bar{a}'\left( \bar{z}_{\epsilon} \right)}.
\end{equation}

$M_n\left( z \right)$, $N_n\left( z \right), a\left( z \right)$ are analytic and bounded for $\left| z \right|>1$ and continuous for $\left| z \right|\ge1$, and $\bar{M}_n\left( z \right)$, $\bar{N}_n\left( z \right), \bar{a}\left( z \right)$ are analytic and bounded for $\left| z \right|<1$ and continuous for $\left| z \right|\le 1$. Moreover, $b\left( z \right)$ and $\bar{b}\left( z \right)$ are continuous in $\left| z \right|=1$.

The asymptotic behaviors of the Jost functions and scattering data with respect to the spectral parameter $z$ are as follows
\begin{equation}
M_n\left( z \right) =\left( \begin{array}{c}
	1+O\left( z^{-2},\ even \right)\\
	z^{-1}Q_{n-1}+O\left( z^{-3},\,\,odd \right)\\
\end{array} \right) ,\ \ N_n\left( z \right) =\left( \begin{array}{c}
	-z^{-1}c_{n}^{-1}Q_n+O\left( z^3,\,\,odd \right)\\
	c_{n}^{-1}+O\left( z ^2,\,\,even \right)\\
\end{array} \right) ,\,\left| z \right|\rightarrow \infty ,
\end{equation}
\begin{equation}
\bar{M}_n\left( z \right) =\left( \begin{array}{c}
	z Q_{n-1}+O\left( z^3,\,\,odd \right)\\
	1+O\left( z^2,\,\,even \right)\\
\end{array} \right) ,\,\,\bar{N}_n\left( z \right) =\left( \begin{array}{c}
	c_{n}^{-1}+O\left( z^2,\,\,even \right)\\
	-zc_{n}^{-1}Q_n+O\left( z^3,\,\,odd \right)\\
\end{array} \right) ,\,\ z \rightarrow 0,
\end{equation}
\begin{subequations}
\begin{align}
a\left( z \right) &=1+O\left( z ^{-2},even \right), \ \ \left| z \right|\rightarrow \infty,\\
\bar{a}\left( z \right) &=1+O\left( z^2,even \right), \,\,\,\,\ \ z \rightarrow 0.
\end{align}
\end{subequations}

When $R_n=\mp Q_n$, there are induced symmetries

$\bullet$ (Jost functions) $\bar{M}_n\left( z \right) =\left( \begin{matrix}
	0&		\mp1\\
	1&		0\\
\end{matrix} \right) M_{n}^{*}\left( \frac{1}{z^*} \right),\ \
\bar{N}_n\left( z \right) =\mp\left( \begin{matrix}
	0&		\mp1\\
	1&		0\\
\end{matrix} \right) N_n^{*}\left( \frac{1}{z^*} \right)$.

$\bullet$ (Scattering matrix) $a\left( -z \right)=a\left( z \right),\ b\left( -z \right)=-b\left( z \right),\ \bar{a}\left( z \right) =a^*\left( \frac{1}{z^*} \right)$,\ \ $\bar{b}\left( z \right) =\mp b^*\left( \frac{1}{z^*} \right)$.

$\bullet$ (Reflection coefficients) $\rho\left( -z \right)=-\rho\left( z \right),\ \bar{\rho}\left( -z \right)=-\bar{\rho}\left( z \right),\ \bar{\rho}\left( z \right) =\mp \rho^*\left( \frac{1}{z^*} \right)$.

$\bullet$ (Norming constants) $\bar{C}_j=\pm\left( z_{j}^{*} \right) ^{-2}C_{j}^{*}$. \\
The symmetry $\bar{a}\left( z \right) =a^*\left( 1/z^* \right)$ implies that $z_j$ is an eigenvalue if, and only if, $\bar{z}_j=1/z_j^*$ is an eigenvalue. Hence, $J=\bar{J}$, namely, $a$ and $\bar{a}$ have the same number of zeros.
\subsection{Time evolution}
\hspace{0.6cm}In the process of direct scattering, we omit the contribution of time, from now on we will write explicitly the time dependence of the scattering data.

Based on the asymptotic behavior of the time spectrum problem subject to boundary conditions and the relationship between the eigenfunctions, the time evolution of the scattering data for Eq. (2.10) is
\begin{equation}
\frac{\partial a\left(z,\tau \right)}{\partial \tau}=0,\  \frac{\partial \bar{a}\left(z,\tau \right)}{\partial \tau} =0,
\end{equation}
\begin{equation}
\frac{\partial \rho\left(z,\tau \right)}{\partial \tau}-\mathscr{N}_1\left( z^2 \right)\rho\left(z,\tau \right)=0,\ \frac{\partial \bar{\rho}\left(z,\tau \right)}{\partial \tau}+\mathscr{N}_1\left( z^2 \right)\bar{\rho}\left(z,\tau \right)=0,
\end{equation}
\begin{equation}
\frac{\partial C_j\left( \tau \right)}{\partial \tau}-\mathscr{N}_1\left( z_j^2 \right) C_j\left(\tau \right)=0,\ \frac{\partial \bar{C_j}\left(\tau \right)}{\partial \tau}+\mathscr{N}_1\left( \bar{z}_j^2 \right)\bar{C_j}\left(\tau \right)=0,
\end{equation}
where $\mathscr{N}_1\left( z^2 \right)=\left(z^{-2}-z^2\right)^{1+\epsilon}$.

Similar arguments can be used to obtain the time dependence of the other fractional integrable discrete mKdV hierarchy.
\subsection{Inverse scattering}
\hspace{0.6cm}In this subsection, we reconstruct the potential by using Gel'fand-Levitan-Marchenko equations and Riemann-Hilbert problem.
\subsubsection{Gel'fand-Levitan-Marchenko equations}
\hspace{0.6cm}By applying the integral factors and triangular kernels to Eqs. (4.6), we get the following GLM equations
\begin{subequations}
\begin{align}
\bar{\kappa}\left( n,m,\tau \right) +\left( \begin{array}{c}
	0\\
	1\\
\end{array} \right) F_R\left( m+n,\tau \right) +\sum_{\underset{j+m=odd}{j=n+1}}^{+\infty}{\kappa \left( n,j,\tau \right) F_R\left( m+j,\tau \right)}=0,\\
\kappa \left( n,m,\tau \right) +\left( \begin{array}{c}
	1\\
	0\\
\end{array} \right) \bar{F}_R\left( m+n,\tau \right) +\sum_{\underset{j+m=odd}{j=n+1}}^{+\infty}{\bar{\kappa}\left( n,j,\tau \right) \bar{F}_R\left( m+j,\tau \right)}=0,
\end{align}
\end{subequations}
where
\begin{subequations}
\begin{align}
F_R\left( n,\tau \right) &=\left\{ \begin{array}{l}
	2\sum_{j=1}^J{z_{j}^{-n-1}C_j\left( \tau \right)}+\frac{1}{\pi i}\int_{C_R}{z^{-n-1}\rho \left( z,\tau \right) dz}\ \ n=odd,\\
	0\,\,\,\,\,\,\,\,\,\,\,\,\,\,\,\,\,\,\,\,\,\,\,\,\,\,\,\,\,\,\,\,\,\,\,\,\,\,\,\,\,\,\,\,\,\,\,\,\,\,\,\,\,\,\,\,\,\,\,\,\,\,\,\,\,\,\,\,\,\,\,\,\,\,\,\,\,\,\,\,\,\,\,\,\,\,\,\,\,\,\, \,\,\,\,\,\,\,\,\,\,\,\,\,\,\,\,\,\,\,\, n=even.
\end{array} \right. \\
\bar{F}_R\left( n,\tau \right) &=\left\{ \begin{array}{l}
	2\sum_{j=1}^{\bar{J}}{\bar{z}_{j}^{n-1}\bar{C}_j\left( \tau \right)}+\frac{1}{\pi i}\int_{C_R}{z^{n-1}\bar{\rho}\left( z,\tau \right) dz}\,\,\,\,n=odd,\\
	0\,\,\,\,\,\,\,\,\,\,\,\,\,\,\,\,\,\,\,\,\,\,\,\,\,\,\,\,\,\,\,\,\,\,\,\,\,\,\,\,\,\,\,\,\,\,\,\,\,\,\,\,\,\,\,\,\,\,\,\,\,\,\,\,\,\,\,\,\,\,\,\,\,\,\,\,\,\,\,\,\,\,\,\,\,\,\,\,\,\,\, \,\,\,\,\,\,\,\,\,\,\,\,\ n=even.
\end{array} \right.
\end{align}
\end{subequations}
and $C_R$ denotes the right half of the unit circle.

Comparing the triangular kernel representations of the eigenfunctions with the asymptotic behaviors of the Jost functions, the reconstruction of the potential is
\begin{equation}
Q_n\left( \tau \right) =-\kappa ^{\left( 1 \right)}\left( n,n+1,\tau \right).
\end{equation}

Using the symmetry between triangular kernels, $\kappa ^{\left( 1 \right)}\left( n,n+1,\tau \right)$ can be abbreviated as
\begin{equation}
\kappa ^{\left( 1 \right)}\left( n,n+1,\tau \right) -\bar{F}_R\left( n,\tau \right) \pm \sum_{j'=n+1}^{+\infty}{\sum_{j''=n+1}^{+\infty}{\kappa ^{\left( 1 \right)}\left( n,j'',\tau \right) \bar{F}_{R}^{*}\left( j'+j'',\tau \right)}}\bar{F}_R\left( j'+m,\tau \right) =0.
\end{equation}
The sources of column vectors $\kappa\left( n,m,\tau \right)$ and $\bar{\kappa}\left( n,m,\tau \right)$ are shown in Ref.~\cite{g45}.
\subsubsection{Riemann-Hilbert problem}
\hspace{0.6cm}The boundary condition for $N_n$ depends on $Q_k$ for all $k\ge1$. However, $Q_n$ is unknown in the inverse problem. To remove this dependence, we introduce the functions
\begin{subequations}
\begin{align}
N_{n}^{+}&=\left( \begin{matrix}
	1&		0\\
	0&		c_n\\
\end{matrix} \right) N_n=\left( \begin{array}{c}
	-z ^{-1}c_{n}^{-1}Q_n\\
	1\\
\end{array} \right) +O\left( z ^{-2} \right) ,\ \ \left| z \right|\rightarrow \infty,\\
\mu _{n}^{+}&=\left( \begin{matrix}
	1&		0\\
	0&		c_n\\
\end{matrix} \right) \mu _n=\left( \begin{array}{c}
	1\\
	z ^{-1}c_nQ_{n-1}\\
\end{array} \right) +O\left( z ^{-2} \right) ,\,\,\,\,\left| z \right|\rightarrow \infty,\\
N_{n}^{-}&=\left( \begin{matrix}
	1&		0\\
	0&		c_n\\
\end{matrix} \right) \bar{N}_n=\left( \begin{array}{c}
	c_{n}^{-1}\\
	-z Q_n\\
\end{array} \right) +O\left( z ^2 \right) ,\,\,\,\,\ \ \ \ \ \ \ z \rightarrow 0,\\
\mu _{n}^{-}&=\left( \begin{matrix}
	1&		0\\
	0&		c_n\\
\end{matrix} \right) \bar{\mu}_n=\left( \begin{array}{c}
	z Q_{n-1}\\
	c_n\\
\end{array} \right) +O\left( z ^2 \right) ,\,\,\,\,\,\,\,\,\ \ \ \ \ z \rightarrow 0.
\end{align}
\end{subequations}
{\bf Proposition 4.1.} By defining the following meromorphic matrices
\begin{equation}
\mathbf{m}_{n}^{+}\left( z,\tau \right) =\left( \begin{matrix}
	\mu _{n}^{+}\left( z,\tau \right) ,&		N_{n}^{+}\\
\end{matrix}\left( z,\tau \right) \right) ,\ \mathbf{m}_{n}^{-}\left( z,\tau \right) =\left( \begin{matrix}
	\mu _{n}^{-}\left( z,\tau \right) ,&		N_{n}^{-}\\
\end{matrix}\left( z,\tau \right) \right),
\end{equation}
the Riemann-Hilbert problem can be proposed as follows.

$\bullet$ Analyticity: $\mathbf{m}_{n}^{+}\left( z,\tau \right)$ is analytic in $\left| z \right|>1$ and $\mathbf{m}_{n}^{-}\left( z,\tau \right)$ is analytic in $\left| z \right|<1$, and they all have simple poles.

$\bullet$ Jump relation:
\begin{equation}
\mathbf{m}_{n}^{+}\left( z,\tau \right) =\mathbf{m}_{n}^{-}\left( z,\tau \right) \left( I-\mathbf{V}_n\left( z,\tau \right) \right) ,\ \ \ \ \left| z \right|=1,
\end{equation}
where the jump matrix $\mathbf{V}_n\left( z,\tau \right) =\left( \begin{matrix}
	\rho \left( z,\tau \right) \bar{\rho}\left( z \right)&		z ^{2n}\bar{\rho}\left( z,\tau \right)\\
	-z ^{-2n}\rho \left( z,\tau \right)&		0\\
\end{matrix} \right) $.

$\bullet$ Asymptotic behaviors:
\begin{equation}
\mathbf{m}_{n}^{\pm}\left( z,\tau \right) \rightarrow I,\ \ \ \ \left| z \right|\rightarrow \infty.
\end{equation}

By using the Cauchy operators and Plemelj's formula, one can derive the solution of the Riemann-Hilbert problem. The solution of the Riemann-Hilbert problem can be written as
\begin{equation}
\mathbf{m}_{n}^{\pm}\left( z,\tau \right) =I\pm \underset{\left| \xi \right|\gtrless 1}{\underset{\xi \rightarrow z}{\lim}}\frac{1}{2\pi i}\oint_{\left| \omega \right|=1}{\frac{\mathbf{m}_{n}^{\mp}\left( \omega,\tau \right) \mathbf{V}_n\left( \omega,\tau \right)}{\omega -\xi}}d\omega.
\end{equation}

Considering the asymptotic behaviors and pole contributions, the solution of the Riemann-Hilbert problem becomes
\begin{subequations}
\begin{align}
N_{n}^{+}\left( z,\tau \right) &=\left( \begin{array}{c}
	0\\
	1\\
\end{array} \right) +\sum_{j=1}^{\bar{J}}{\bar{C}_j\left( \tau \right)}\bar{z}_{j}^{2n}\left[ \frac{1}{z -\bar{z}_j}N_{n}^{-}\left( \bar{z}_j,\tau \right) +\frac{1}{z +\bar{z}_j}N_{n}^{-}\left( -\bar{z}_j,\tau \right) \right] + \nonumber\\ &\ \ \ \
\underset{\left| \xi \right|>1}{\underset{\xi \rightarrow z}{\lim}}\frac{1}{2\pi i}\oint_{\left| \omega \right|=1}{\frac{\omega ^{2n}\bar{\rho}\left( \omega,\tau \right) N_{n}^{-}\left( \omega,\tau \right)}{\omega -\xi}}d\omega,\\
N_{n}^{-}\left( z,\tau \right) &=\left( \begin{array}{c}
	1\\
	0\\
\end{array} \right) +\sum_{j=1}^J{C_j\left( \tau \right)}z _{j}^{-2n}\left[ \frac{1}{z -z _j}N_{n}^{+}\left( z _j,\tau \right) +\frac{1}{z +z _j}N_{n}^{+}\left( -z _j,\tau \right) \right] -\nonumber \\ &\ \ \ \
\underset{\left| \xi \right|<1}{\underset{\xi \rightarrow z}{\lim}}\frac{1}{2\pi i}\oint_{\left| \omega \right|=1}{\frac{\omega ^{-2n}\rho \left( \omega,\tau \right) N_{n}^{+}\left( \omega,\tau \right)}{\omega -\xi}d\omega}.
\end{align}
\end{subequations}

By comparing the power series expansion of the right-hand side of Eq. (4.27a) with the expansion (4.22c), we obtain
\begin{equation}
Q_n\left( \tau \right)=2\sum_{j=1}^J{C_j\left( \tau \right)}z _{j}^{-2\left( n+1 \right)}N_{n}^{+\left( 2 \right)}\left( z _j,\tau \right) +\frac{1}{2\pi i}\oint_{\left| \omega \right|=1}{\omega ^{-2\left( n+1 \right)}\rho \left( \omega,\tau \right) N_{n}^{+\left( 2 \right)}\left( \omega,\tau \right) d\omega}.
\end{equation}

\subsection{Fractional soliton solution}
\hspace{0.6cm}Case 1(Fractional one-soliton): Fractional one-soliton solution means that there are one quartet of eigenvalues $\left\{ \pm z_1,\ \pm \bar{z}_1 \right\}$ with $\left| z_1 \right|>1$ and $\left| \bar{z}_1 \right|<1$, by choosing appropriate parameters, fractional one-soliton solution can be obtained with two methods of RH problem and GLM equations as shown in Figs. 1 and 2.
\begin{figure}[H]
\centering
 \begin{minipage}[c]{0.3\textwidth} 
  \centering
  \includegraphics[width=\textwidth]{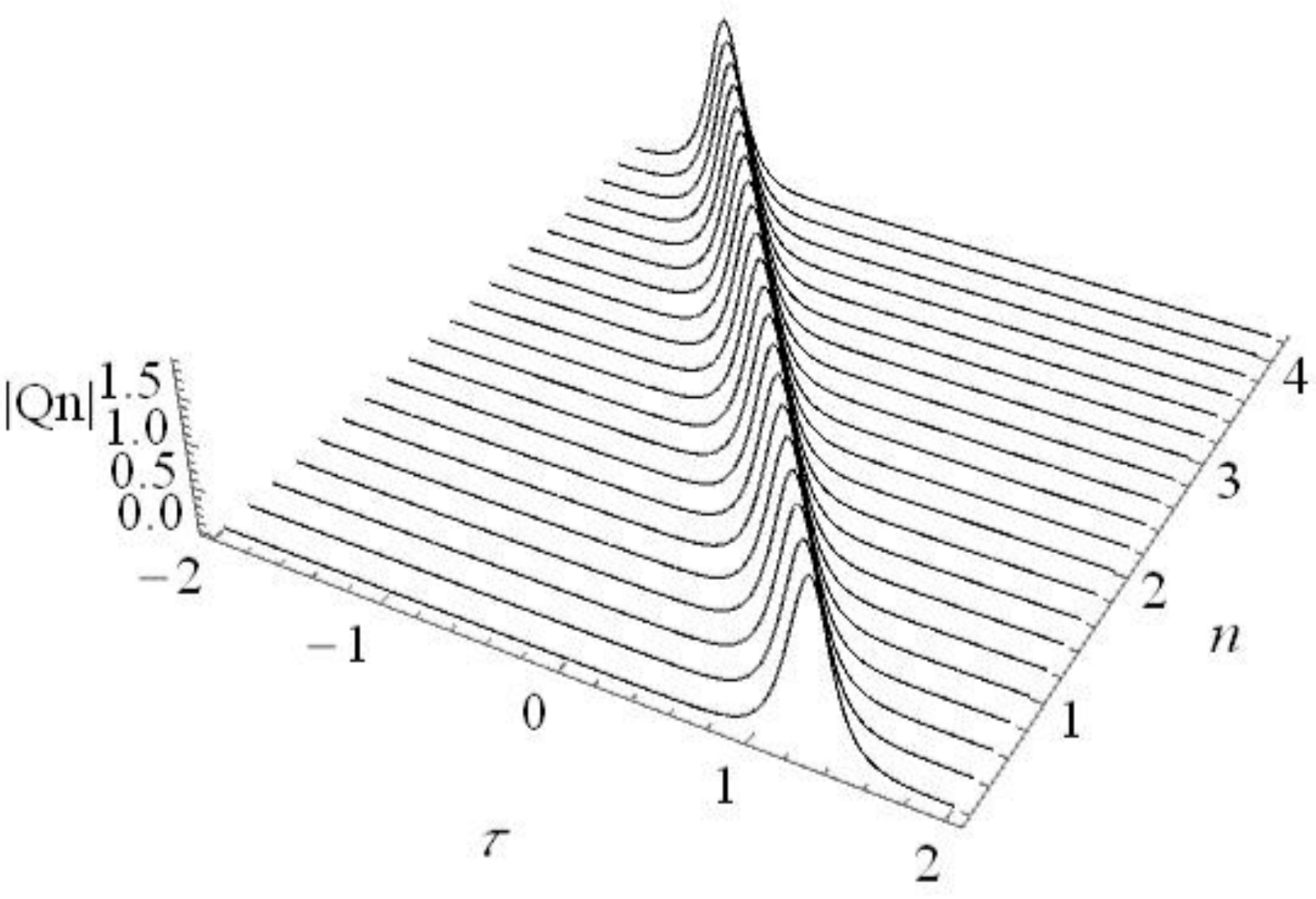} 
  \centerline{(a)}
 \end{minipage}
 \begin{minipage}[c]{0.3\textwidth}
  \centering
  \includegraphics[width=\textwidth]{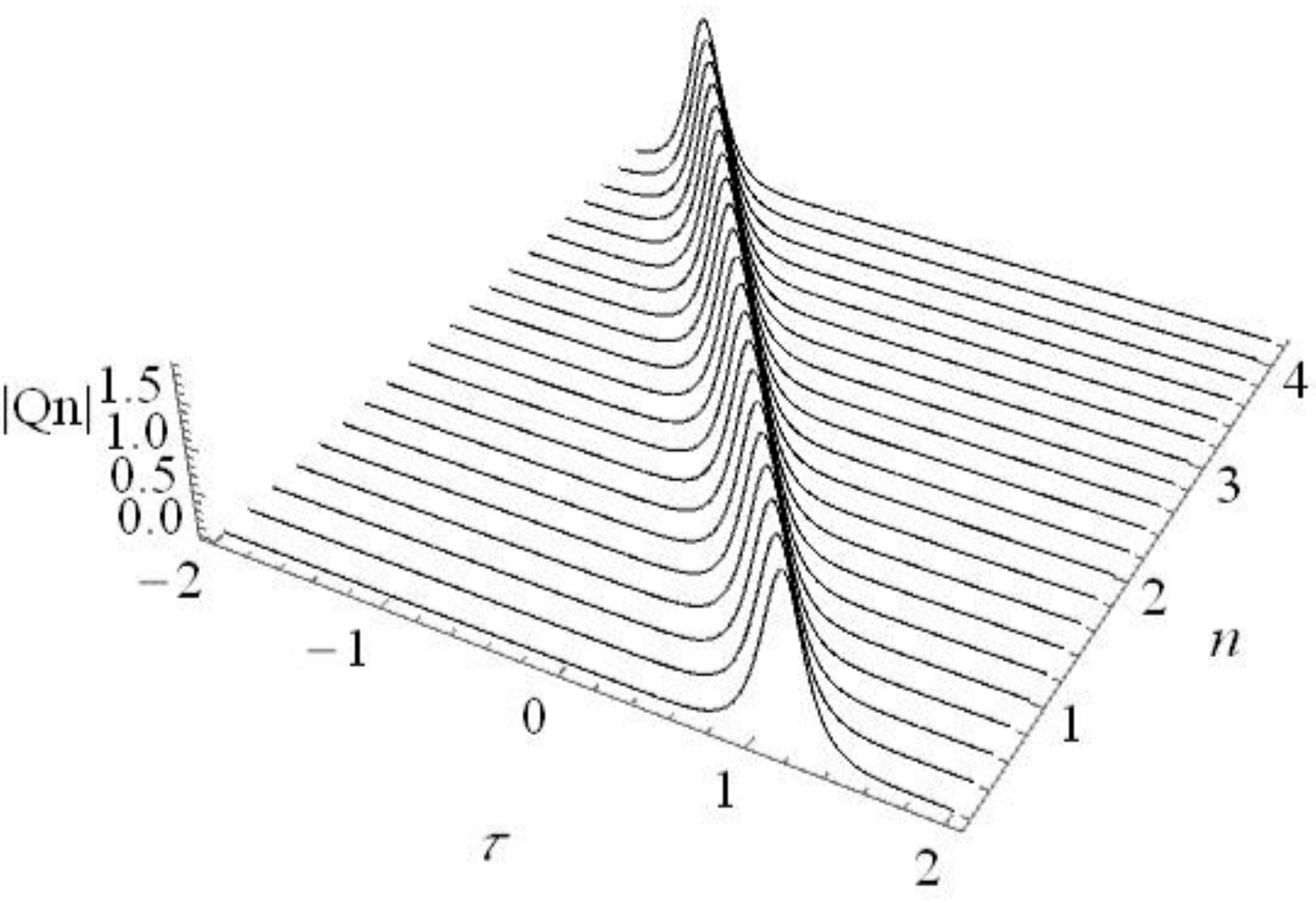}
  \centerline{(b)}
 \end{minipage}
\caption{Fractional one-soliton solution of Eq. (2.10) with $z_1=2,\ \bar{z}_1=0.5,\ C_1\left(0\right)=\bar{C}_1\left(0\right)=1,\ \epsilon=0.25$. \ (a) GLM equations; (b) RH problem.}
\end{figure}

\begin{figure}[H]
\centering
 \begin{minipage}[c]{0.3\textwidth} 
  \centering
  \includegraphics[width=\textwidth]{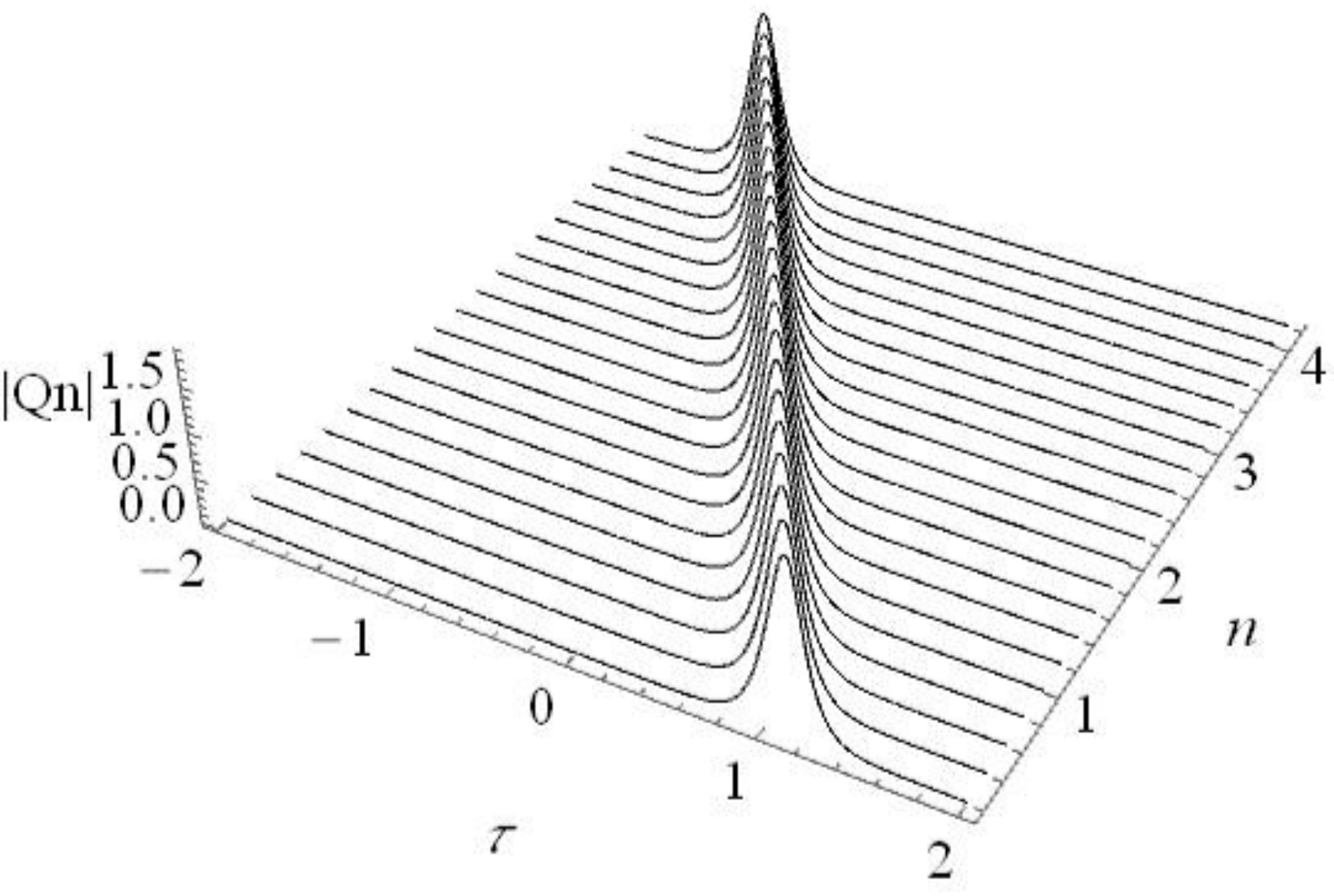} 
  \centerline{(a)}
 \end{minipage}
 \begin{minipage}[c]{0.3\textwidth}
  \centering
  \includegraphics[width=\textwidth]{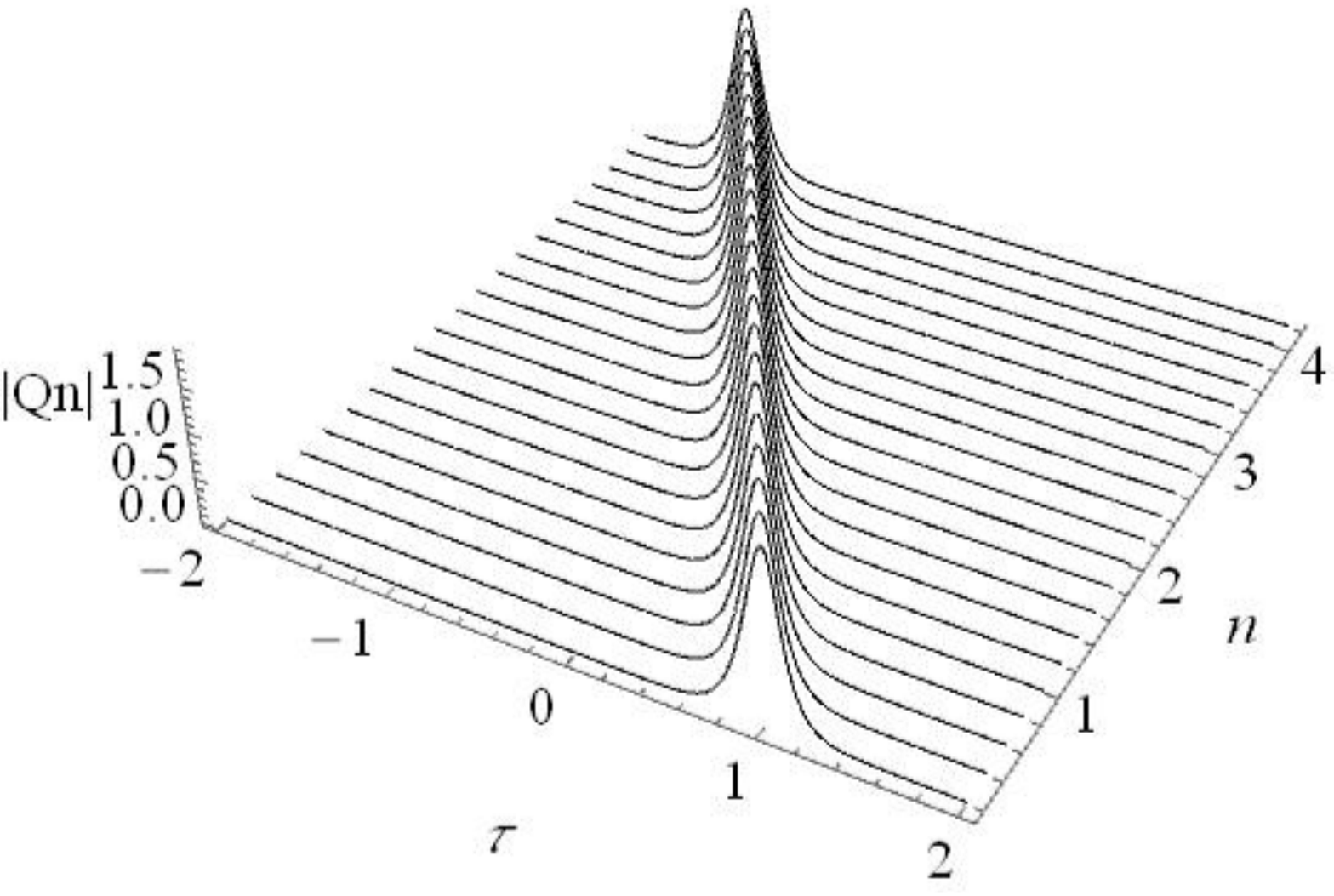}
  \centerline{(b)}
 \end{minipage}
\caption{Fractional one-soliton solution of Eq. (2.11) with $z_1=2,\ \bar{z}_1=0.5,\ C_1\left(0\right)=\bar{C}_1\left(0\right)=1,\ \epsilon=0.25$. \ (a) GLM equations; (b) RH problem.}
\end{figure}

Case 2(Fractional two-soliton): Fractional two-soliton solution has double poles $\left\{ \pm z_1,\ \pm \bar{z}_1,\ \pm z_2,\ \pm \bar{z}_2 \right\}$.  A similar method holds for two-soliton solution as shown in Figs. 3 and 4.
\begin{figure}[H]
\centering
  \includegraphics[width=0.5\textwidth]{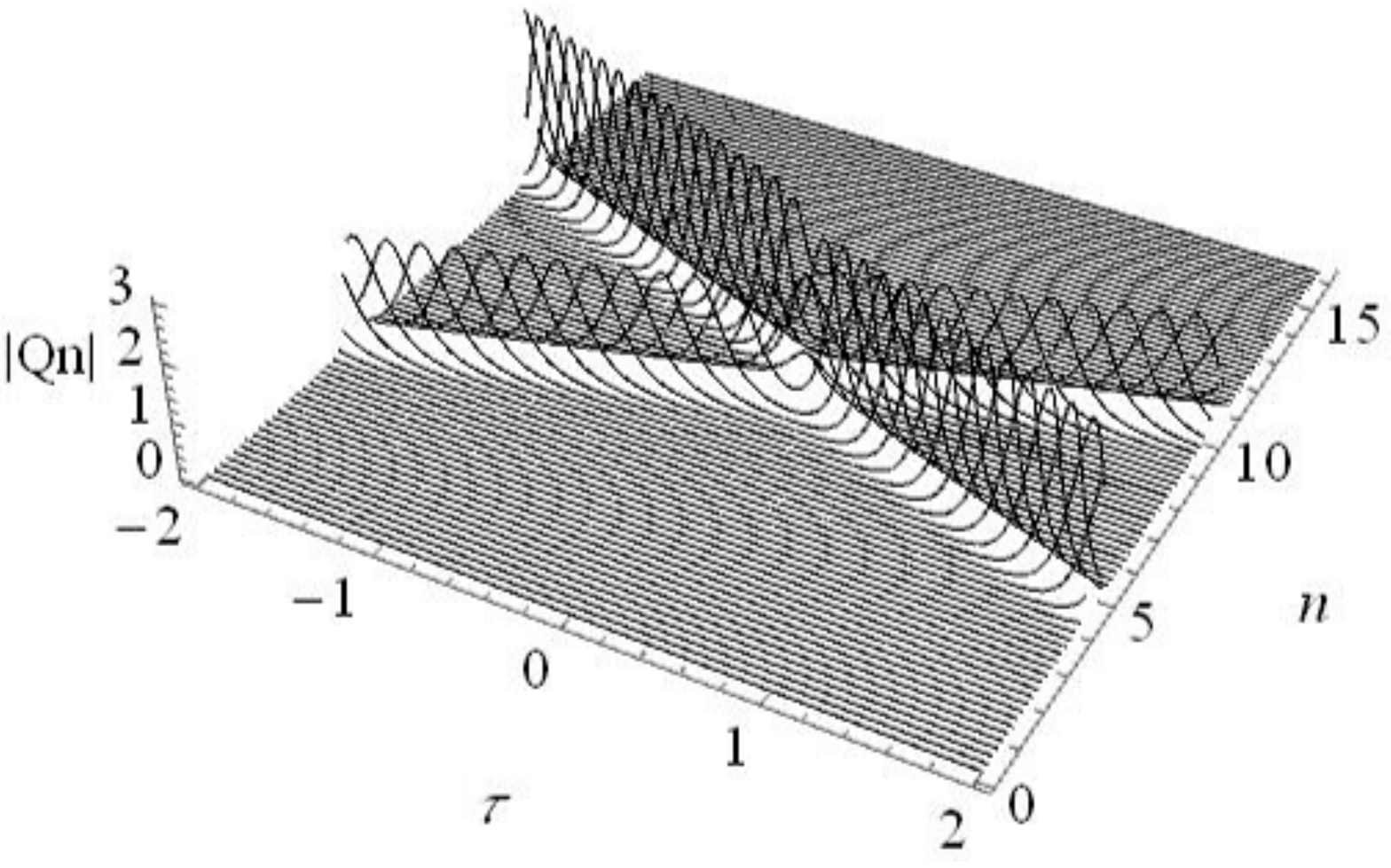} 
\caption{Fractional two-soliton solution of Eq. (2.10) with $z_1=2.5,\ \bar{z}_1=0.4,\ z_2=2,\ \bar{z}_2=0.5, \ C_1\left(0\right)=\bar{C}_1\left(0\right)=C_2\left(0\right)=\bar{C}_2\left(0\right)=1,\ \epsilon=0.1$.}
\end{figure}
\begin{figure}[H]
\centering
  \includegraphics[width=0.5\textwidth]{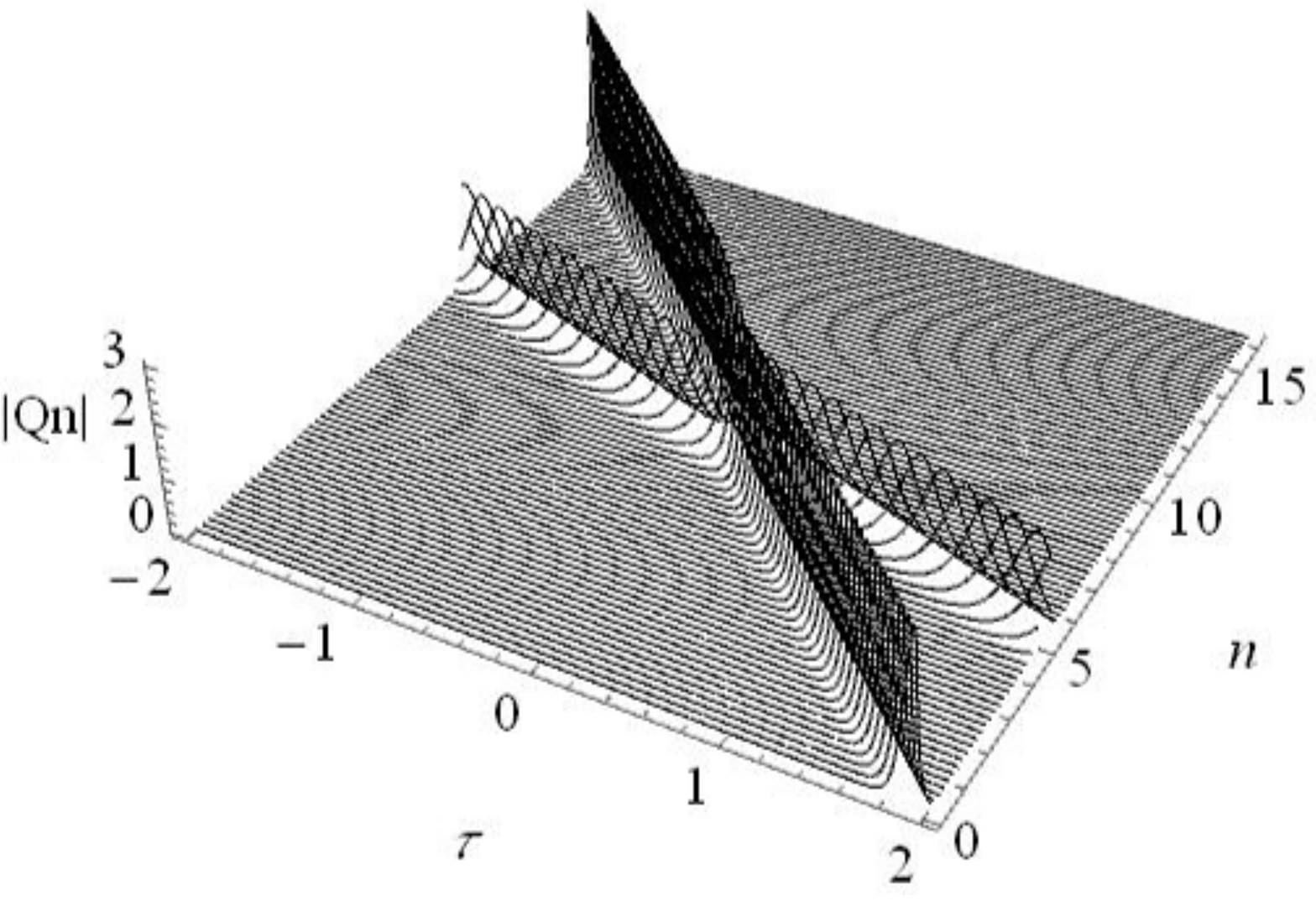} 
\caption{Fractional two-soliton solution of Eq. (2.11) with $z_1=2.5,\ \bar{z}_1=0.4,\ z_2=2,\ \bar{z}_2=0.5, \ C_1\left(0\right)=\bar{C}_1\left(0\right)=C_2\left(0\right)=\bar{C}_2\left(0\right)=1,\ \epsilon=0.1$.}
\end{figure}

\section{Solitary wave solution of the fAdmKdV equation}
\hspace{0.6cm}In this section we obtain the numerical solutions of the fAdmKdV equation, which is another fractional discrete form of Eq.$\ $(2.5), unlike Eq.$\ $(2.10), which arises by changing the discrete central difference operator into a discrete fractional central difference operator in the following form
\begin{equation}
Q_{n,\tau}-\left( Q_{n+1}-Q_{n-1} \right) ^{1+\epsilon}\mp Q_{n}^{2}\left( Q_{n+1}-Q_{n-1} \right) =0.
\end{equation}

The fAdmKdV equation is solved numerically by a split-step Fourier scheme, which decomposes the original problem into linear and nonlinear subproblems which take into account purely dispersive and purely nonlinear effects. The linear subproblem is treated accurately, and the nonlinear subproblem is approximately solved by the fourth-order Runge-Kutta scheme. The initial condition is given by the soliton solution to Eq.$\ $(2.10) using the IST
\begin{equation}
Q_n\left( \tau \right) =\sinh \left( 2\eta h \right) \text{sech} \left( 2\eta h\pm \omega _1\left( z_1 \right) \tau+\phi _0 \right),
\end{equation}
where $\omega _1\left( z_1 \right) =\left(z_{1}^{-2}-z_{1}^{2}\right)^{1+\epsilon}, \phi _0=-\ln \left( h\left| \bar{C}_1\left( 0 \right) \right|/\sinh\left( 2h\eta \right) \right), z_1=\exp\left( h\eta \right)$. The discrete Fourier transform of linear subproblem is
\begin{equation}
\left( Q_{n+1}-Q_{n-1} \right) ^{1+\epsilon}=\frac{1}{2\pi}\int_{-\pi /h}^{\pi /h}{dk\hat{Q}\left( k \right) e^{iknh}\left( -2i\sin kh \right) ^{1+\epsilon}},
\end{equation}
where $h$ is the distance between lattice sites and $\hat{Q}\left( k \right) =h\sum_{n=-\infty}^{\infty}{Q_ne^{-iknh}}$ is the Fourier transform of $Q_n$. We adopt the second-order split-step Fourier scheme
\begin{equation}
Q_n\left( \tau_{m+1} \right) =\exp \left( \frac{1}{2}\Delta \tau_m\mathcal{V} \right) \exp \left( \Delta \tau_m\mathcal{U} \right) \exp \left( \frac{1}{2}\Delta \tau_m\mathcal{V} \right),
\end{equation}
where $\mathcal{U}Q_n=\left( Q_{n+1}-Q_{n-1} \right) ^{1+\epsilon}$ and $\mathcal{V}Q_n=\pm Q_{n}^{2}\left( Q_{n+1}-Q_{n-1} \right)$. Throughout this article, results of the fAdmKdV equation are obtained using $N=1000$ spatial grid points and a time step of $\Delta \tau=1\times10^{-3}$.

Fig. 5 shows the solitary wave solutions of the fAdmKdV equation with an amplitude of $1.175$ and different $\epsilon$ at different simulation times. Fig. 6 shows the solitary wave solutions of the fAdmKdV equation with different amplitudes of $1.1752, 1.9043$ and same $\epsilon=0.1$ at different simulation times. Recall that amplitude is $A=\sinh\left( 2\eta h \right)$ which is related to the parameters $\eta$ and $h$ ($h$ is taken to be $1$). In order to better observe that the peaks of the numerical solution decrease with the increase of time, we have panned the spatial locations where the numerical solution peaks are located as far as possible into the same region. As the amplitude increases and $\epsilon$ increases, the decay rate of the solitary wave solution of the fAdmKdV equation becomes faster.
\begin{figure}[H]
\centering
 \begin{minipage}[c]{0.45\textwidth} 
  \centering
  \includegraphics[width=\textwidth]{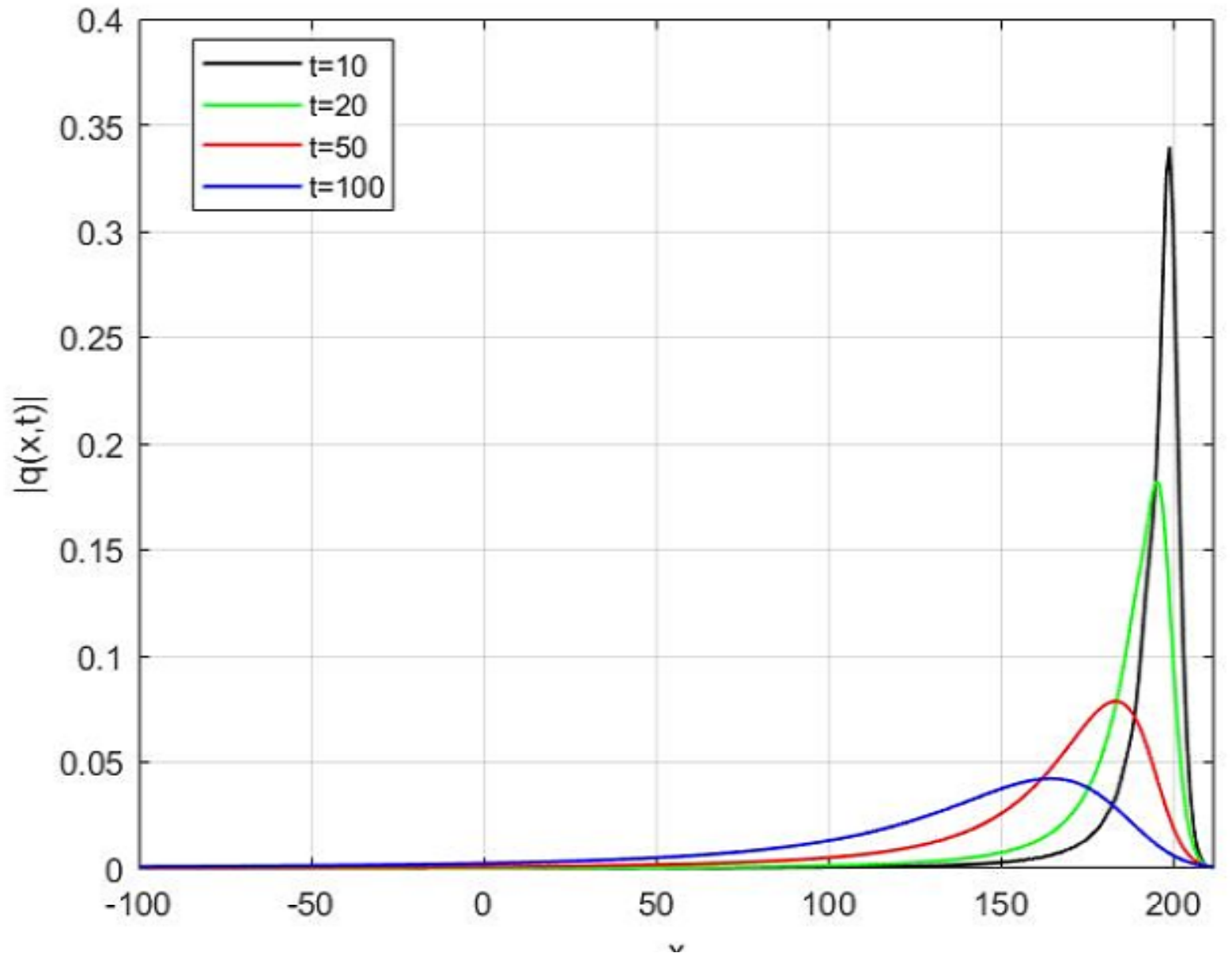} 
  \centerline{(a)}
 \end{minipage}
 \begin{minipage}[c]{0.45\textwidth}
  \centering
  \includegraphics[width=\textwidth]{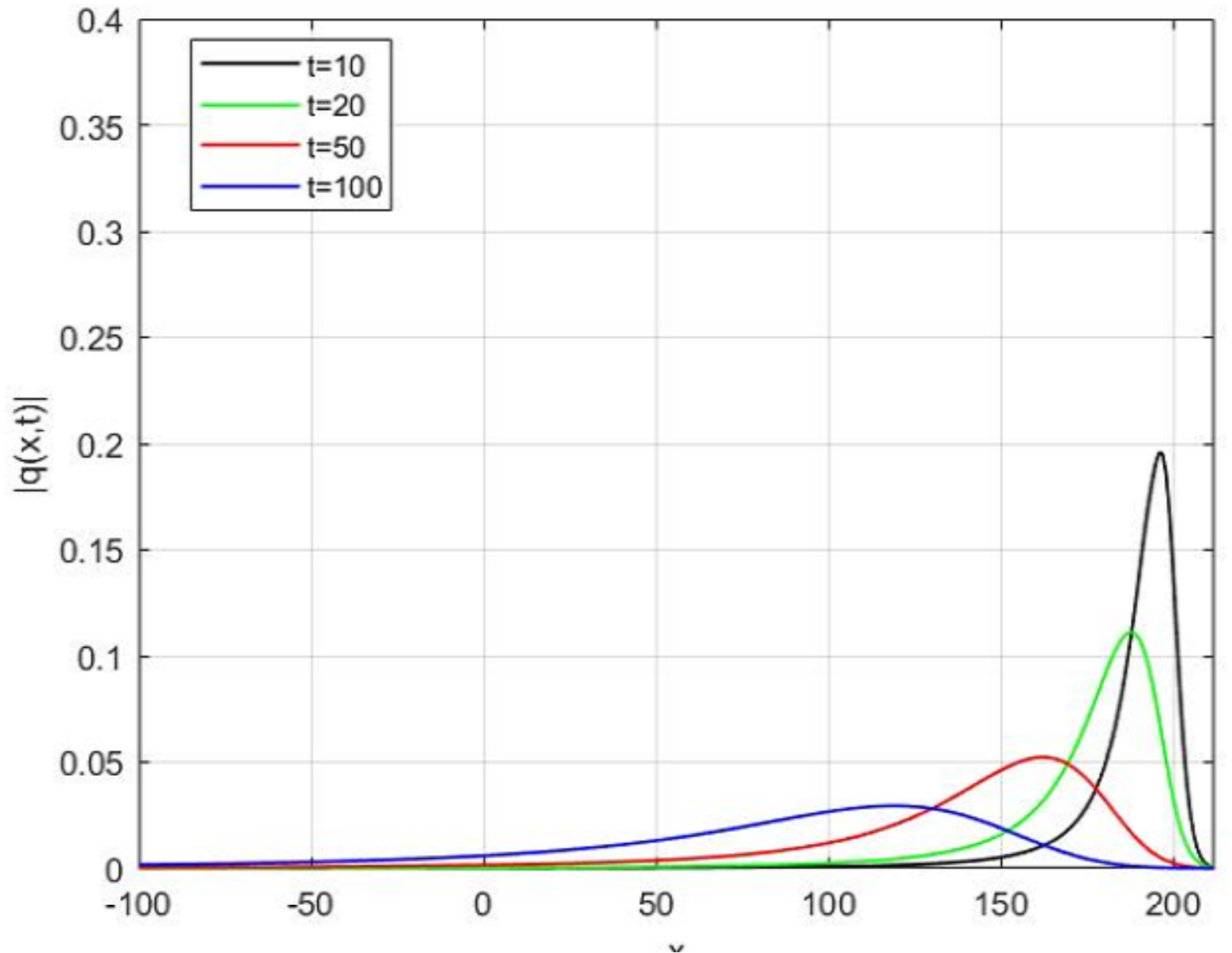}
  \centerline{(b)}
 \end{minipage}
\caption{The solitary wave solutions of the fAdmKdV equation at different simulation times with $h=1,\ \eta=0.5,\ \phi_0=0$. \ (a) $\epsilon=0.1$; (b) $\epsilon=0.2$.}
\end{figure}
\begin{figure}[H]
\centering
 \begin{minipage}[c]{0.45\textwidth} 
  \centering
  \includegraphics[width=\textwidth]{51.pdf} 
  \centerline{(a)}
 \end{minipage}
 \begin{minipage}[c]{0.45\textwidth}
  \centering
  \includegraphics[width=\textwidth]{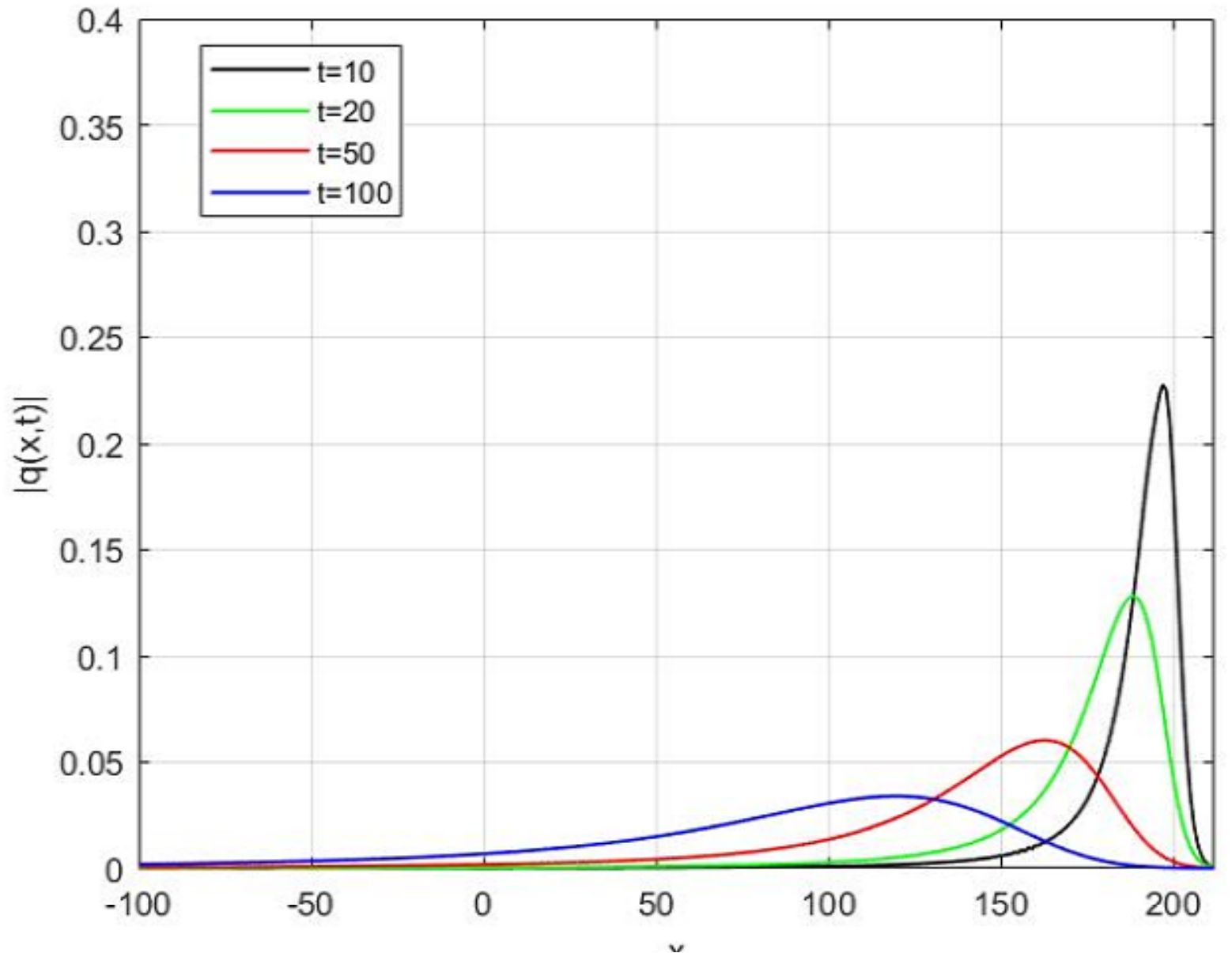}
  \centerline{(b)}
 \end{minipage}
\caption{The solitary wave solutions of the fAdmKdV equation at different simulation times with $h=1,\ \epsilon=0.1,\ \phi_0=0$. \ (a) $\eta=0.5$; (b) $\eta=0.6$.}
\end{figure}
\begin{figure}[H]
\centering
  \includegraphics[width=0.5\textwidth]{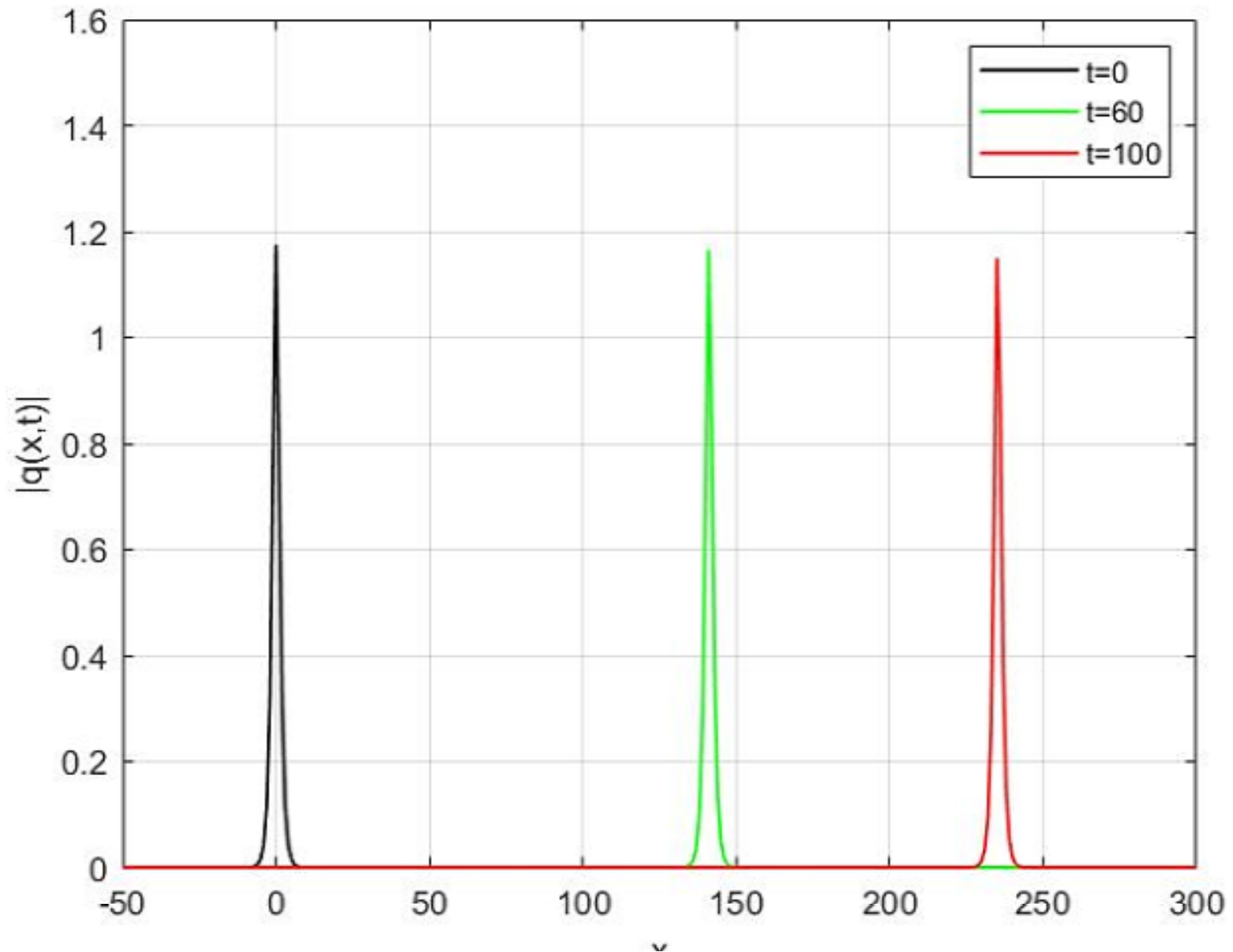} 
\caption{The soliton solutions of the discrete mKdV equation at different simulation times with $h=1,\ \epsilon=0,\ \eta=0.5,\ \phi_0=0$.}
\end{figure}

Fig. 7 exhibits the soliton solutions of the discrete mKdV equation with an amplitude of $1.175$ at different simulation times, i.e., the situation of the fAdmKdV equation when $\epsilon=0$. This is a standard soliton solution using the same second-order split-step Fourier scheme as the above figures. We calculate the averaged peak amplitude of the soliton solution in Fig. 7, as shown in Fig. 8, which is a straight line parallel to the time axis, indicating that the amplitude settles down to a constant. The averaged peak amplitude is obtained by taking the mean of the first $10$ points and the last $10$ points around each point as the value of the point.
\begin{figure}[H]
\centering
  \includegraphics[width=0.5\textwidth]{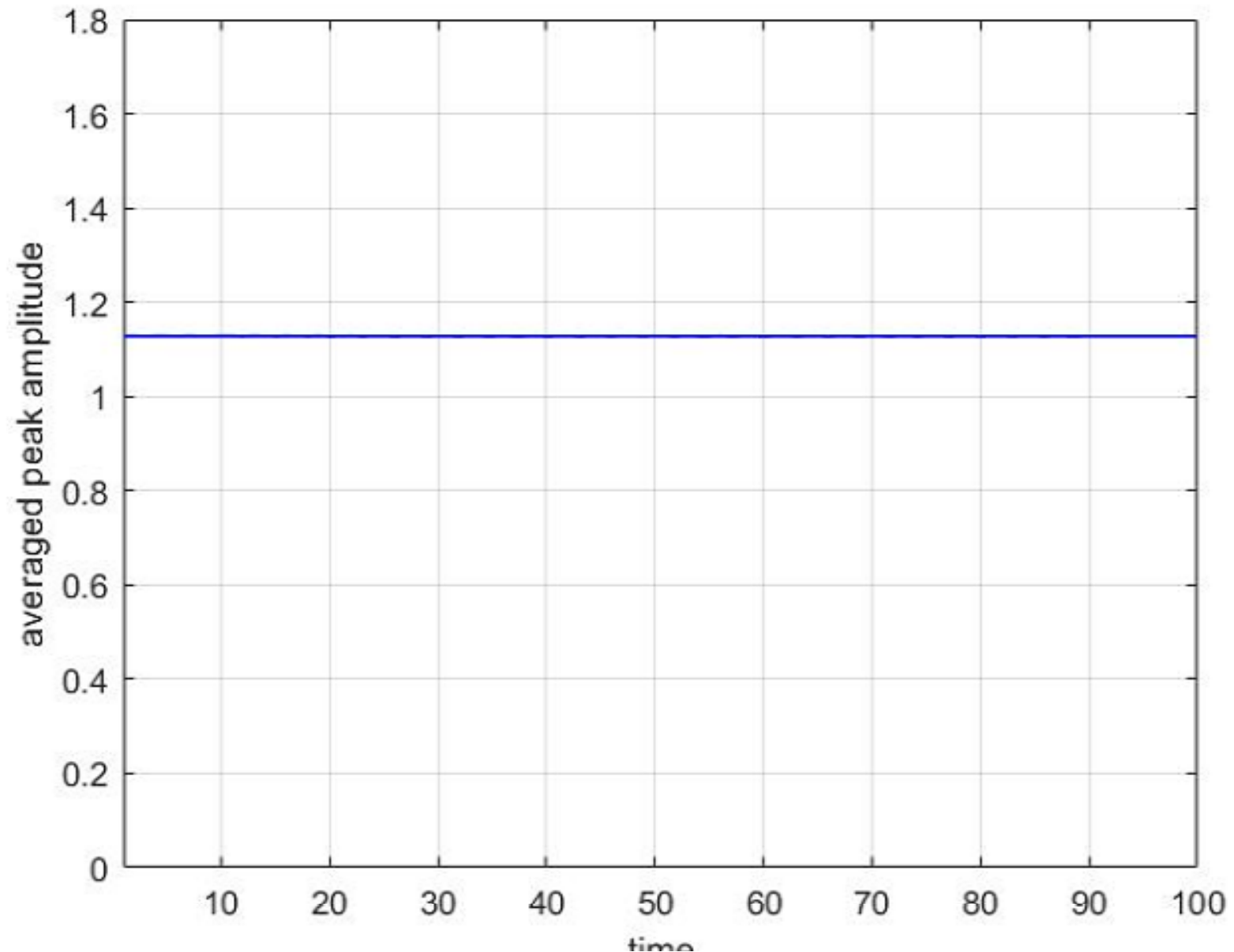} 
\caption{The time-averaged amplitudes of the soliton solutions of the discrete mKdV equation with $h=1,\ \epsilon=0,\ \eta=0.5,\ \phi_0=0$.}
\end{figure}
Fig. 9 displays the density diagram of the solitary wave solution of the fAdmKdV equation with the amplitude of $1.175$ and $\epsilon=0.1$. Throughout the manuscript, the initial situation of numerical simulation selection is that, putting $\tau=0$ into Eq. (5.2).
\begin{figure}[H]
\centering
  \includegraphics[width=0.5\textwidth]{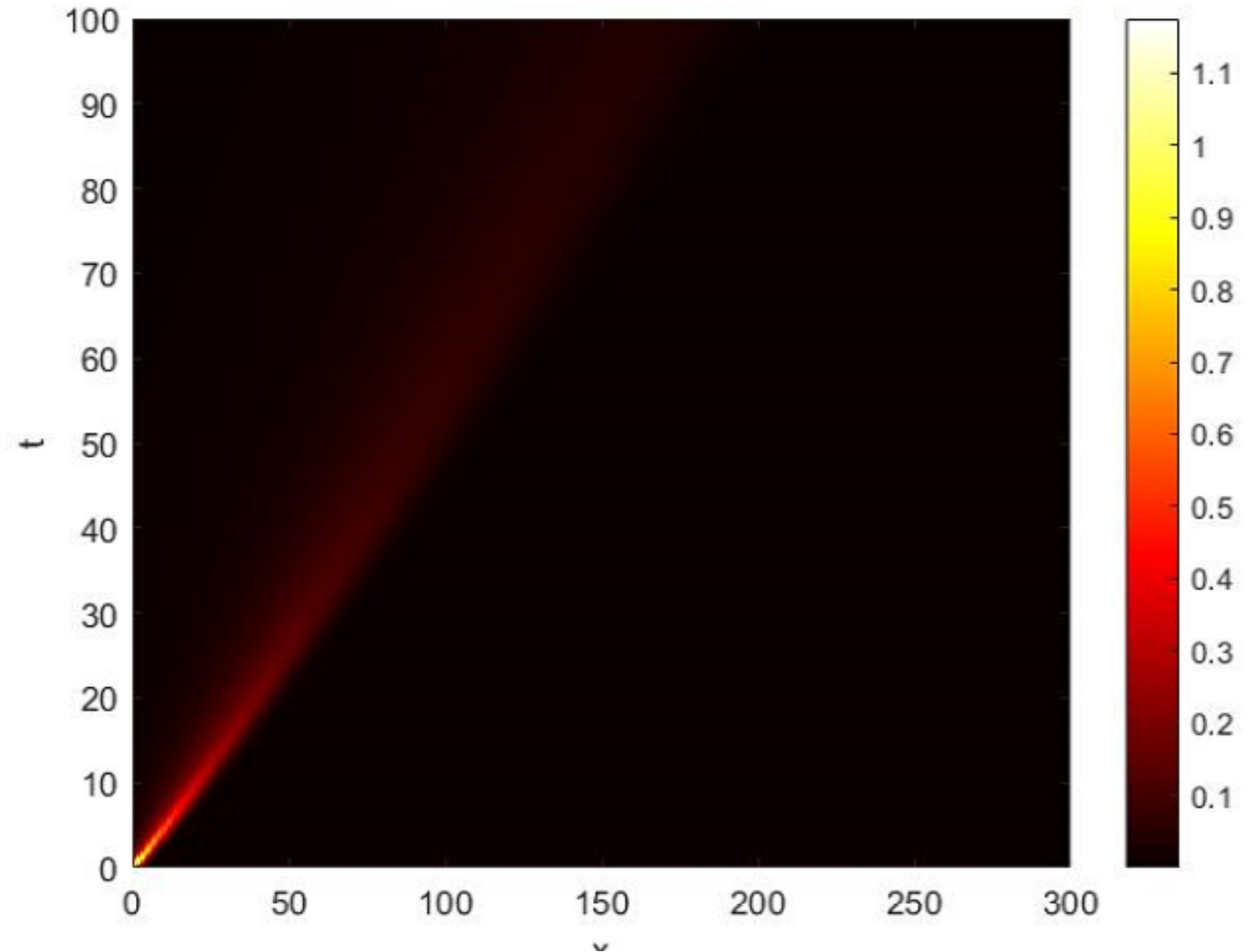} 
\caption{The density diagram of solitary wave solution of the fAdmKdV equation with $h=1,\ \epsilon=0.1,\ \eta=0.5,\ \phi_0=0$.}
\end{figure}
The fractional one-soliton velocity of Eq. (2.10) is
\begin{equation}
\vartheta =-\frac{2\sinh^{1+\epsilon}\left( -2\eta h \right)}{2\eta h}.
\end{equation}
Fig. 10 reveals the variation of soliton velocity for small $(\eta=0.05)$, medium $(\eta=0.1)$ and large $(\eta=0.5)$ amplitudes with the change of fractional parameter $\epsilon\in[0, 1]$. Compared with the fmKdV and fsineG equations, the relationship between the velocity and amplitude of soliton in Fig. 10 is much more complicated.
\begin{figure}[H]
\centering
  \includegraphics[width=0.5\textwidth]{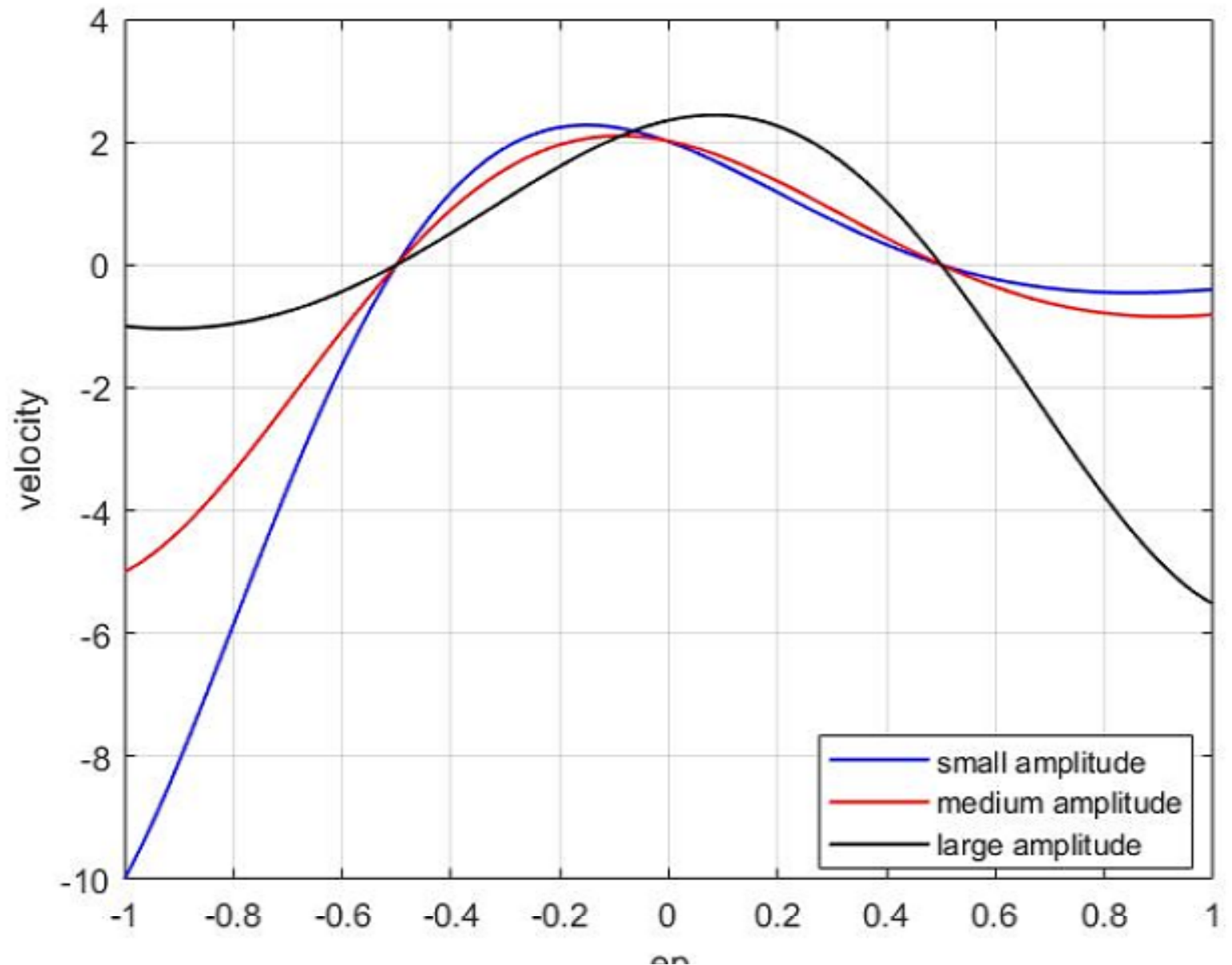} 
\caption{Velocity of the fractional one-soliton solution to the Eq. (2.10) for small$(A=0.1002)$, medium$(A=0.2013)$ and large$(A=1.1752)$ amplitudes with $h=1$.}
\end{figure}

\section{Conclusions}
\hspace{0.7cm}In conclusion, the fractional integrable discrete modified Korteweg-de Vries hierarchy was obtained. We have derived the fractional integrable discrete modified Korteweg-de Vries equations through three key components: linear dispersion relations, completeness relations, and IST. Particularly, we have presented a rigorous theory of the IST and thus have given their single-poles solutions for the reflectionless potentials by solving the corresponding GLM equations and RH problem. Velocity of the fractional one-soliton solution to the fractional integrable discrete modified Korteweg-de Vries equations has been gotten, which is more complicated than its continuous counterpart. The numerical solutions of the non-integrable fractional averaged discrete modified Korteweg-de Vries equation have been obtained by using the second-order split-step Fourier scheme. In addition, various properties of those solutions have been discussed by virtue of graphic simulation. This work can be extended to other fractional integrable discrete nonlinear evolution hierarchies, and to fully discrete systems.

\vspace{5mm}\noindent\textbf{Acknowledgements}
\\\hspace*{\parindent}We express our sincere thanks to each member of our discussion group for their suggestions. This work has been supported by the National Natural Science Foundation of China under Grant No. 11905155, and the Fund Program for the Scientific Activities of Selected Returned Overseas Scholars in Shanxi Province under Grant No. 20220008.

\vspace{5mm}\noindent\textbf{Data Availability Statement}
\\\hspace*{\parindent}No data associated in the manuscript.


\begin{thebibliography}{99}

\bibitem{g1}
P.\ G.\ Drazin. {\em Nonlinear Systems}.\ {Cambridge: Cambridge University Press},\ 1992.

\bibitem{g2}
D.\ K.\ Arrowsmith, C.\ M.\ Place. {\em An introduction to dymamical systems}.\ {Cambridge: Cambridge University Press},\ 1990.

\bibitem{g3}
S.\ Wiggins. {\em Introduction to applied nonlinear dyanmical systems and chaos}.\ {New York: Springer-Verlag},\ 1990.

\bibitem{g4}
M.\ A.\ Helal. {Soliton solution of some nonlinear partial differential equations and its applications in fluid mechanics}.\ {\em Chaos Solitons Fractals}\ {\bf 13(9)} (2002) 1917-1929.

\bibitem{g5}
G.\ Wysin, A.\ R.\ Bishop, P.\ Kumar. {Soliton dynamics on an easy-plane ferromagnetic chain}.\ {\em J. Phys. C: Solid State Phys.}\ {\bf 17(33)} (1984) 5975-5991.

\bibitem{g6}
E.\ A.\ Kuznetsov, A.\ M.\ Rubenchik, V.\ E.\ Zakharov. {Soliton stability in plasmas and hydrodynamics}.\ {\em Phys. Rep.}\ {\bf 142(3)} (1986) 103-165.

\bibitem{g7}
Z.\ Y.\ Sun, Y.\ T.\ Gao, X.\ Yu, Y.\ Liu. {Amplification of nonautonomous solitons in the Bose-Einstein condensates and nonlinear optics}.\ {\em EPL}\ {\bf 93(4)} (2011) 40004.

\bibitem{g8}
M.\ J.\ Ablowitz, P.\ A.\ Clarkson. {\em Solitons, nonlinear evolution equations and inverse scattering}.\ {Cambridge: Cambridge University Press},\ 1991.

\bibitem{g9}
S.\ W.\ Xu, J.\ S.\ He, L.\ H.\ Wang. {The Darboux transformation of the derivative nonlinear Schr\"{o}dinger equation}.\ {\em J. Phys. A: Math. Theor.}\ {\bf 44(30)} (2011) 305203.

\bibitem{g10}
C.\ J.\ Wang. {Lump solution and integrability for the associated Hirota bilinear equation}.\ {\em Nonlinear Dyn.}\ {\bf 87(4)} (2017) 2635-2642.

\bibitem{g11}
J.\ Zhang, H.\ Q.\ Hao. {Soliton solutions of the AB system via the Jacobi elliptic function expansion method}.\ {\em Optik}\ {\bf 244} (2021) 167541.

\bibitem{g12}
C.\ S.\ Gardner, J.\ M.\ Greene, M.\ D.\ Kruskal, R.\ M.\ Miura. {Method for solving Korteweg-de Vries equation}.\ {\em Phys. Rev. Lett.}\ {\bf 19(19)} (1967) 1095-1097.

\bibitem{g13}
M.\ J.\ Ablowitz, D.\ J.\ Kaup, A.\ C.\ Newell, H.\ Segur. {The inverse scattering transform-Fourier analysis for nonlinear problems}.\ {\em Stud. Appl. Math.}\ {\bf 53(4)} (1974) 249-315.

\bibitem{g14}
K.\ B.\ Oldham, J.\ Spanier. {\em The Fractional Calculus}.\ {New York-London: Academic Press},\ 1974.

\bibitem{g15}
I.\ Podlubny. {Fractional differential equations}.\ {\em Math. Sci. Eng.}\ {\bf 198} (1999) 41-119.

\bibitem{g16}
B.\ Ross. {A brief history and exposition of the fundamental theory of fractional calculus}.\ {\em Lect. Notes Math.}\ {\bf 457(1)} (1975) 1-36.

\bibitem{g17}
N.\ Laskin. {Fractional quantum mechanics and L$\acute{e}$vy path integrals}.\ {\em Phys. Lett. A}\ {\bf 268(4-6)} (2000) 298-305.

\bibitem{g18}
S.\ B.\ Yuste, K.\ Lindenberg. {Subdiffusion-Limited A+A reactions}.\ {\em Phys. Rev. Lett.}\ {\bf 87(11)} (2001) 118-301.

\bibitem{g19}
R.\ Metzler, J.\ Klafter. {The random walk's guide to anomalous diffusion: a fractional dynamics approach}.\ {\em Phys. Rep.}\ {\bf 339(1)} (2000) 1-77.

\bibitem{g20}
I.\ M.\ Sokolov, A.\ V.\ Chechkin. {Anomalous diffusion and generalized diffusion equations}.\ {\em Fluct. Noise Lett.}\ {\bf 5(2)} (2005) 275-282.

\bibitem{g21}
E.\ K.\ Lenzi, A.\ Somer, R.\ S.\ Zola, L.\ R.\ da Silva, M.\ K.\ Lenzi. {A generalized diffusion equation: solutions and anomalous diffusion}.\ {\em Fluids}\ {\bf 8(34)} (2023) 34.

\bibitem{g22}
Y.\ H.\ Gong, W.\ Yan, M.\ Luo, C.\ X.\ Wang, L.\ Zhang. {A frequency-tunable fractional-N PLL for high-energy physics experiments}.\ {\em IEEE Trans. Nucl. Sci.}\ {\bf 70(4)} (2023) 722-729.

\bibitem{g23}
K.\ B.\ Wu, L.\ Wei, Z.\ X.\ Wang. {Anomalous diffusion and generalized diffusion equations}.\ {\em Plasma Sci. Technol.}\ {\bf 24(4)} (2022) 109-116.

\bibitem{g24}
B.\ Z.\ Han, D.\ S.\ Yin, Y.\ F.\ Gao. {Analysis of the variable-order fractional viscoelastic modeling with application to polymer materials}.\ {\em Polym. Adv. Technol.}\ {\bf 34(8)} (2023) 2707-2720.

\bibitem{g25}
J.\ F.\ Han, C.\ P.\ Li, S.\ D.\ Zeng. {Applications of generalized fractional hemivariational inequalities in solid viscoelastic contact mechanics}.\ {\em Commun. Nonlinear Sci. Numer. Simul.}\ {\bf 115} (2022) 106718.

\bibitem{g26}
Y.\ Y.\ Zhuang, X.\ Song. {Towards a better understanding of fractional brownian motion and its application to finance}.\ {\em Bull. Malaysian Math. Sci. Soc.}\ {\bf 46(5)} (2023) 150.

\bibitem{g27}
M.\ J.\ Ablowitz, J.\ B.\ Been, L.\ D.\ Carr. {Fractional integrable nonlinear soliton equations}.\ {\em Phys. Rev. Lett.}\ {\bf 128(18)} (2022) 184101.

\bibitem{g28}
M.\ J.\ Ablowitz, J.\ B.\ Been, L.\ D.\ Carr. {Integrable fractional modified Korteweg-de Vries, sine-Gordon, and sinh-Gordon equations}.\ {\em J. Phys. A}\ {\bf 55(38)} (2022) 384010.

\bibitem{g29}
M.\ H.\ Zhang, W.\ F.\ Weng, Z.\ Y.\ Yan. {Interactions of fractional $N$-solitons with anomalous dispersions for the integrable combined fractional higher-order mKdV hierarchy}.\ {\em Phys. D}\ {\bf 444} (2023) 133614.

\bibitem{g30}
W.\ F.\ Weng, M.\ H.\ Zhang, G.\ Q.\ Zhang, Z.\ Y.\ Yan. {Dynamics of fractional $N$-soliton solutions with anomalous dispersions of integrable fractional higher-order nonlinear Schr\"{o}dinger equations}.\ {\em Chaos}\ {\bf 32(12)} (2022) 123110.

\bibitem{g31}
L.\ An, L.\ M.\ Ling, X.\ E.\ Zhang. {Nondegenerate solitons in the integrable fractional coupled Hirota equation}.\ {\em Phys. Lett. A}\ {\bf 460} (2023) 128629.

\bibitem{g32}
D.\ S.\ Mou, C.\ Q.\ Dai, Y.\ Y.\ Wang. {Integrable fractional $n$-component coupled nonlinear Schr\"{o}dinger model and fractional $n$-soliton dynamics}.\ {\em Chaos Solitons Fractals}\ {\bf 171} (2023) 113451.

\bibitem{g33}
L.\ An, L.\ M.\ Ling, X.\ E.\ Zhang. {Inverse scattering transform for the integrable fractional derivative nonlinear Schr\"{o}dinger equation}.\ {\em Phys. D}\ {\bf 458} (2024) 133888.

\bibitem{g34}
L.\ An, L.\ M.\ Ling. {The Riemann-Hilbert approach for the integrable fractional Fokas-Lenells equation}.\ {\em arXiv preprint arXiv:2308.16640,}\ 2023.

\bibitem{g35}
M.\ J.\ Ablowitz, J.\ B.\ Been, L.\ D.\ Carr. {Fractional integrable and related discrete nonlinear Schr\"{o}dinger equations}.\ {\em Phys. Lett. A}\ {\bf 452} (2022) 128459.

\bibitem{g36}
D.\ Tavares, R.\ Almeida, D.\ F.\ M.\ Torres. {Caputo derivatives of fractional variable order: Numerical approximations}.\ {\em Commun. Nonlinear Sci. Numer. Simul.}\ {\bf 35} (2016) 69-87.

\bibitem{g37}
T.\ R.\ Taha, M.\ J.\ Ablowitz. {Analytical and numerical aspects of certain nonlinear evolution equations II. Numerical, nonlinear Schr\"{o}dinger equation}.\ {\em J. Comput. Phys.}\ {\bf 55(2)} (1984) 203-230.

\bibitem{g38}
T.\ R.\ Taha, M.\ J.\ Ablowitz. {Analytical and numerical aspects of certain nonlinear evolution equations. III. Numerical, Korteweg-deVries equation}.\ {\em J. Comput. Phys.}\ {\bf 55(2)} (1984) 231-253.

\bibitem{g39}
T.\ R.\ Taha, M.\ J.\ Ablowitz. {Analytical and numerical aspects of certain nonlinear evolution equations IV. Numerical, modified Korteweg-de Vries equation}.\ {\em J. Comput. Phys.}\ {\bf 77(2)} (1988) 540-548.

\bibitem{g40}
G.\ M.\ Muslu, H.\ A.\ Erbay. {A split-step Fourier method for the complex modified Korteweg-de Vries equation}.\ {\em Comput. Math. with Appl.}\ {\bf 45(1)} (2003) 503-514.

\bibitem{g41}
G.\ M.\ Muslu, H.\ A.\ Erbay. {Higher-order split-step Fourier schemes for the generalized nonlinear Schr\"{o}dinger equation}.\ {\em Math. Comput. Simul.}\ {\bf 67(6)} (2004) 581-595.

\bibitem{g42}
C.\ Han, Y.\ L.\ Wang, Z.\ Y,\ Li. {Numerical solutions of space fractional variable-coefficient KdV-modified KdV equation by fourier spectral method}.\ {\em Fractals}\ {\bf 29(8)} (2021) 1-19.

\bibitem{g43}
V.\ Gerdjikov, M.\ Ivanov, P.\ Kulish. {Expansions over the ``squared'' solutions and difference evolution equations}.\ {\em J. Math. Phys.}\ {\bf 25(1)} (1984) 25-34.

\bibitem{g44}
S.\ -C.\ Chiu, J.\ F.\ Ladik. {Generating exactly soluble nonlinear discrete evolution equations by a generalized Wronskian technique}.\ {\em J. Math. Phys.}\ {\bf 18(4)} (1977) 690-700.

\bibitem{g45}
M.\ J.\ Ablowitz, B.\ Prinari, A.\ D.\ Trubatch. {\em Discrete and Continuous Nonlinear Schr\"{o}dinger Systems}.\ {Cambridge: Cambridge University Press},\ 2004.
\end{thebibliography}
\end{document}